\documentclass{aa}

\usepackage{graphics}

\begin{document}

\thesaurus{03(11.01.2; 11.06.2; 11.14.1; 11.16.1)}

\title{The optical properties of low redshift radio galaxies}

\author{Federica Govoni \inst{1, 3}, Renato Falomo \inst{1}, Giovanni Fasano \inst{1} and Riccardo Scarpa \inst{2}}
\offprints{R. Falomo (falomo@pd.astro.it)\\
Based on observations collected at the European Southern Observatory,
La Silla, Chile.\\
Based on observations collected at the Nordic Optical Telescope, La Palma.}

\institute{Osservatorio Astronomico di Padova, Vicolo dell'Osservatorio 5, I--35122 
Padova, Italy
\and
Space Telescope Science Institute, 3700 San Martin Drive, Baltimore, MD 
21218, U.S.A
\and
Dipartimento di Astronomia dell'Universit\'a di Bologna, Via Ranzani 1, I--40127 Bologna, Italy.}

\date{Received; Accepted}

\maketitle

\markboth{F. Govoni et al.: The optical properties of low redshift radio galaxies }{F. Govoni et al.: The optical properties of low redshift radio galaxies}

\begin{abstract}

We present morphological and photometric properties of 79 low
redshift ($z\leq0.12$) radio galaxies extracted from two radio flux
limited samples of radio sources.

All objects are imaged in the R band and for a subsample we have also
obtained B band images.  The sample includes sources of both FRI and
FRII radio morphological type. Through the decomposition of the
luminosity profiles and the analysis of the structural profiles
(ellipticity, PA, c4) of the galaxies we are able to characterize in
detail the optical properties of the radio galaxies.

It is found that most of host galaxies are luminous bulge dominated
systems similar to normal giant ellipticals.  Some cases of additional
disk components are found whose spheroid-to-disk luminosity ratio is
similar to that found in S0 galaxies.

The average absolute magnitude is $\langle M_{HOST}(tot)
\rangle=-24.0$ with a clear trend for FRI sources to be $\sim0.5$ mag
brighter than FRII galaxies.

In about 40\% of the objects observed we find an excess of light in
the nucleus that is attributed to the presence of a nuclear point
source whose luminosity is on average $\sim$ 1-2\% of the total flux
of the host galaxy.  The luminosity of these nuclear point sources
appears correlated with the core radio power of the galaxies.

Radio galaxies follow the same $\mu_e$ - R$_e$ relationship as normal
 elliptical galaxies.
 
The distribution of ellipticity, the amount of twisting and shape of
isophotes ($boxy , disky$) do not differ significantly from other
ellipticals. The evidence for recent interactions is therefore rather
modest.

Finally on average radio galaxies are bluer and have
a color dispersion larger than normal elliptical galaxies, and
the average color gradient in radio galaxies appears slightly
steeper than in normal ellipticals.

These results support a scenario where radio emission is
weakly related with the overall properties and/or the activity have
negligible effects on the global characteristics of the host galaxy.



\end{abstract}

\section{Introduction}

Radio galaxies are the nearest kind of radio-loud active galactic nuclei
(AGN). These objects can be detected and studied in the local universe as
well as at very large distances (Pentericci et al. 1999).

Advances in radio astronomy 
have produced a wealth of data and information on the properties of the 
radio emitting regions at various physical scales.
A simple but important morphological classification of extended radio sources 
was made by Fanaroff \& Riley (1974), who pointed out that low-power 
sources tend to be brighter close to their nuclei whereas high-power 
ones are brighter at their outer extremities.
Although on the kpc scale
 radio galaxies of high and
low luminosity have quite different radio morphologies, VLBI observations
 suggest that both FR I and II sources have similar
relativistic flows on parsec scale (Giovannini et al. 1994).

In the optical band photometric studies of radio galaxies
 have been carried out by various authors for different
 samples of radio galaxies
(Hine \& Longair 1979; Longair \&
Seldner 1979; Lilly \& Prestage 1987; Prestage \& Peacock 1988; Owen
\& Laing 1989; Smith \& Heckman 1989a,b;
 Owen \& White 1991; Gonzalez-Serrano et al. 1993;
de Juan et al. 1994; Colina \& de Juan 1995; Ledlow \& Owen 1995 ).

Optical studies have been focused either on the morphological properties
of host galaxies or on the properties
of their close
environment. Moreover many of these investigations have considered specific
 kinds of sources: powerful radio galaxies ( Lilly \& Prestage 1987,
Smith \& Heckman 1989b);
 FRI radio sources (Colina \& de Juan 1995);
 radio galaxies in clusters, mostly FRI sources (Ledlow \& Owen 1995).

Depending on the quality of data and the kind of analysis
 performed it was possible to extract both photometric and
 morphological properties of galaxies: total optical luminosity, overall
 morphology,
 scale length (effective radius), ellipticity, isophote twisting,
 presence of nuclear components, centering of isophotes,
 isophotal shape, presence of close
 companions and statistics about disturbed morphology.

Usually, host galaxies of radio sources 
are found to be luminous and large ellipticals often interacting with
close companion galaxies.
Distortion of the isophotes (displacement of the center and twisting),
excesses over a de Vaucouleurs law,
and the presence of nearby companions  have been reported 
in an high fraction of FRI galaxies (Colina \& de Juan 1995) and
is assumed to be
related with strong interactions/collisions between galaxies.
Deviation of isophotes from pure ellipses 
were reported by
Bender et al. (1987) and interpreted as an indication of recent merging.
Smith \& Heckman (1989b), looking at a sample of 72
powerful radio galaxies (at 178 MHz
$P_{178}\geq 5\times 10^{24}$ Watt/Hz), concluded that
galaxy interactions/mergers play an important role
in the powerful radio galaxy phenomenon. They found
that over 50\% of the sample galaxies exhibited peculiar optical morphologies
such as shells, tidal bridges, and distorted isophotes. 

The scenario which is emerging from the above mentioned studies is
mainly focused on the connection between the radio activity phenomenon
and the occurrence of close encounters and merging events in different
environments.
In particular, galaxies hosting FRII radio sources
could result from interactions, in relatively poor environments,
involving at least one gas--rich, dynamically cold galaxy (i.e. a disk
galaxy). Instead, gravitational interactions and merging in
richer environments, involving early-type galaxies alone, could give
rise to FRI radio sources (see e.g. Colina \& Perez-Fournon 1990).

In the light of this scenario,
an important question that should be addressed is whether the ambient medium can trigger 
the radio emission. Fanti (1984), was the first to suggest that the probability of radio
emission from elliptical galaxies does not depend on their location inside or outside
the cluster. This was confirmed by a study on a large sample of radio galaxies 
by Ledlow \& Owen (1995), who also conclude that the properties of radio
sources do not
depend on the optical richness of the cluster within which the galaxy resides. This would 
imply that the probability for a galaxy to form a radio source is not dependent on the
environment. The probability of detecting radio emission from an elliptical 
galaxy is mostly dependent on the optical luminosity.  
Ledlow \& Owen (1995) have also compared  the optical
properties of a sample of 265  FRI radio galaxies in
rich clusters with a sample of
radio-quiet galaxies selected from the same environment. They
found that the local density of nearby companions and the frequency of
morphological peculiarities or tidal interactions are not significantly
different between the radio-loud and radio-quiet samples.
Therefore the link between 
radio activity and galaxy interaction is not yet clear and
the point must be further investigated.


\begin{table*}
\begin{center}
\begin{tabular}{cccccccc}
\multicolumn{8}{c}{{\bf Table 1.} Radio properties}\\
\multicolumn{8}{c}{}\\
\hline\\
\multicolumn{1}{c}{IAU name}
&\multicolumn{1}{c}{z} &\multicolumn{1}{c}{Smp.} &\multicolumn{1}{c}{FR-class}
 & \multicolumn{1}{c}{Ref.$^a$} & \multicolumn{1}{c}{$S_{408MHz}$(Jy)} & \multicolumn{1}{c}{$S_{2.7GHz}$(Jy)} & \multicolumn{1}{c}{$S(core)_{4.8GHz}$(Jy)} \\
\medskip  (1)     & (2)	& (3) & (4)& (5) & (6)& (7)& (8) \\
\hline\\
$0005-199$& 0.121&EK&    I &   2  &     2.08  &    0.45&   0.014 \\ 
$0013-316$& 0.107&EK&    I &   2  &     1.36  &    0.24&   0.005 \\
$0023-333$& 0.05 &EK&    I &   2  &     4.09  &    0.73&   0.010 \\
$0034-014$& 0.073&WP&  I/II&   1  &     9.74  &    2.56&   0.299 \\
$0055-016$& 0.045&WP&    I &   1  &    10.9   &    3.46&   0.093 \\
$0123-016$& 0.018&WP&  I/II&   1  &    16.4   &    3.29&   0.100 \\
$0131-367$& 0.03 &WP&    II&   1,2,3&   ...   &    5.6 &   0.038 \\
$0229-208$& 0.089&EK&    I &   2  &     2.52  &    0.75&   0.093 \\
$0247-207$& 0.087&EK&    I &   2  &     3.3   &    0.54&   0.023 \\
$0255+058$& 0.023&WP&    I &   1  &    16.2   &    3.3 &   0.039 \\
$0257-398$& 0.066&EK&    II&   2  &     2.57  &    0.61&   $<$0.004 \\
$0307-305$& 0.066&EK&    II&   2  &	 1.98  &   0.64&   0.003 \\
$0312-343$& 0.067&EK&    I &   2  &     0.94  &    0.25&   0.025 \\
$0325+023$& 0.03 &WP&    II&   1  &    10.90  &    3.18&   0.159 \\
$0332-391$& 0.063&EK&	 I &   2,3&	 4.2  &    0.94&   0.013 \\
$0344-345$& 0.053&EK&    I &   2,3&     7.6   &    2.08&   0.040 \\
$0349-278$& 0.066&EK&   II &   1,2&    13.7   &    3.24&   0.016 \\
$0427-539$& 0.038&WP&    I &   1  &    14.6   &    3.84&   0.057 \\
$0430+052$& 0.033&WP&    I &   1  &     6.08  &    3.0 &   3.458 \\
$0434-225$& 0.069&EK&    I &   2  &     2.26  &    0.66&   0.009 \\
$0446-206$& 0.073&EK&    I &   2  &     2.3   &    0.38&   0.009 \\
$0449-175$& 0.031&EK&    I &   2  &     1.85  &    0.64&   0.010 \\
$0452-190$& 0.039&EK&    I &   2  &     0.66  &    0.25&   0.025 \\
$0453-206$& 0.035&WP&  I/II&   1  &    11.3   &    2.79&   0.04  \\
$0511-305$& 0.058&EK&    II&   2,3&     7.8   &    1.66&   0.010 \\
$0533-377$& 0.096&EK&    I &   2  &     0.96  &    0.3 &   0.013 \\ 
$0546-329$& 0.037&EK&    I &   2,3&     2.89  &    0.43&   0.032 \\
$0548-317$& 0.034&EK&    II&   2  &     2.3   &    0.68&   $<$0.006 \\
$0620-526$& 0.051&WP&    I &   1  &     9.3   &    2.1 &   0.260 \\
$0625-354$& 0.055&WP&    I &   2  &     7.83  &    2.9 &   0.600 \\
$0625-536$& 0.054&WP&    II&   1  &    20.2   &    3.7 &   0.042 \\
$0634-205$& 0.056&EK&  I/II&   2  &    22.8   &    4.5 &   0.012 \\
$0712-349$& 0.044&EK&    I &   2  &     0.74  &    0.21&   0.025 \\
$0718-340$& 0.029&EK&  I/II&   2  &     4.1   &    1.36&   0.030 \\  
$0806-103$& 0.11 &WP&   II &   1  &    13.7   &    2.49&   0.055 \\
$0915-118$& 0.054&WP&    I &   1  &   132.0   &   23.5 &   0.217 \\
$0940-304$& 0.038&EK&    I &   2  &     0.68  &    0.29&   0.047 \\
$0945+076$& 0.086&WP&   II &   1  &    22.1   &    4.3 &   0.032 \\
$1002-320$& 0.089&EK&    I &   2,3&     2.66  &    0.4 &   $<$0.005 \\
$1043-290$& 0.06 &EK&    I &   2  &     0.88  &    0.39&   0.055 \\
$1053-282$& 0.061&EK&    I &   2  &     3.51  &    1.28&   0.115 \\ 
$1056-360$& 0.07 &EK&  I/II&   2,3&     4.09  &    0.96&   0.050 \\
$1107-372$& 0.01 &EK&    I & 2    &     0.96  &    0.41&   0.010 \\
$1123-351$& 0.032&EK&    I & 2    &     6.61  &    1.43&   0.070 \\
$1251-122$& 0.015&WP&    I & 1    &     14.7  &    4.5 &   0.088 \\
$1251-289$& 0.057&EK&    I & 2    &     4.22  &    0.48&   $<$0.010 \\
$1257-253$& 0.065&EK&    I & 2    &     1.5   &    0.45&   0.016 \\
$1258-321$& 0.017&EK&    I & 2    &     2.99  &    0.92&   0.100 \\
$1318-434$& 0.011&WP&    I & 1    &     ...   &    3.09&   0.58  \\
$1323-271$& 0.044&EK&    U & 2    &     3.75  &    0.94&   0.015 \\
$1333-337$& 0.013&WP&  I/II& 1,2,3&    34.0   &   10.06&   0.297 \\
$1344-241$& 0.02 &EK&    I & 2    &     2.38  &    0.39&   $<$0.004 \\
$1354-251$& 0.038&EK&    I & 2    &     1.08  &    0.43&   0.005 \\
\hline\\
\end{tabular}\\
\end{center}
\end{table*}

\begin{table*}
\begin{center}
\begin{tabular}{cccccccc}
\multicolumn{8}{c}{{\bf Table 1 (continued).} Radio properties}\\
\multicolumn{8}{c}{}\\
\hline\\
\multicolumn{1}{c}{IAU name}
&\multicolumn{1}{c}{z} &\multicolumn{1}{c}{Smp.} &\multicolumn{1}{c}{FR-class} & \multicolumn{1}{c}{Ref.$^a$} & \multicolumn{1}{c}{$S_{408MHz}$(Jy)} & \multicolumn{1}{c}{$S_{2.7GHz}$(Jy)}  & \multicolumn{1}{c}{$S(core)_{4.8GHz}$(Jy)} \\
\medskip  (1)     & (2)	& (3) & (4)& (5) & (6)& (7)& (8) \\
\hline\\
$1400-337$& 0.014&EK&    I & 2    &     6.5   &    0.46&   0.015 \\
$1404-267$& 0.022&EK&    I?& 2    &     1.09  &    0.5   &    0.280  \\
$1514+072$& 0.035&WP&    I & 1    &    25.2   &    2.2   &    0.391  \\
$1521-300$& 0.02 &EK&    I?& 2    &     0.26  &    0.31  &    0.160  \\
$1637-771$& 0.041&WP&   II & 1,3  &    11.4   &    3.77  &    0.184  \\
$1717-009$& 0.031&WP&   II & 1    &   138.0   &   33.8   &    0.117  \\
$1733-565$& 0.098&WP&   II & 1    &    13.0   &    5.2   &    0.68   \\
$1928-340$& 0.098&EK&    I & 2    &     3.52  &    0.36  &    0.021  \\
$1929-397$& 0.073&EK&   II & 2    &     6.41  &    1.54  &    0.014  \\
$1949+023$& 0.059&WP&   II & 1    &    13.6   &    3.68  &    0.01   \\
$1954-552$& 0.058&WP&    I & 1,3  &    14.0   &    3.74  &    0.050  \\
$2013-308$& 0.088&EK&    I &   2,3&     2.47  &    0.50  &    0.010  \\
$2031-359$& 0.088&EK&    I &   2  &     4.4   &    0.93  &    0.012  \\ 
$2040-267$& 0.041&EK&   II &   2  &     5.44   &    1.56 &    0.032   \\
$2058-282$& 0.039&WP&  I/II&   1  &    15.9    &    3.10 &    0.063   \\
$2059-311$& 0.039&EK&    I &   2  &     0.67  &    0.24  &    0.018  \\
$2104-256$& 0.038&WP&  I/II&   1  &    31.0   &    7.3   &    0.058  \\
$2128-388$& 0.018&EK&    I &   2  &     1.79  &    0.65  &    0.020  \\
$2158-380$& 0.034&EK&   II &   2,3&     4.12  &    1.01  &    0.005  \\
$2209-255$& 0.063&EK&    I?&   2  &     0.5   &    0.26  &    0.050  \\
$2221-023$& 0.057&WP&   II &   1  &    17.5   &    3.46  &    0.086  \\  
$2225-308$& 0.056&EK&    I &   2,3&     2.07  &    0.55  &    0.030  \\
$2236-176$& 0.07 &EK&    I &   2  &     4.31  &    1.06  &    0.010  \\
$2333-327$& 0.052&EK&   II &   2  &     0.67  &    0.23  &    0.012  \\
$2350-375$& 0.116&EK&    I &   2  &     0.76  &    0.25  &    0.004  \\
$2353-184$& 0.073&EK&    I &   2  &     2.05  &    0.49  &    0.010  \\
\hline\\
\multicolumn{8}{l}{\scriptsize $^a$ References :}\\
\multicolumn{8}{l}{\scriptsize (1) Morganti et al. (1993) or Morganti et al. (1997);}\\
\multicolumn{8}{l}{\scriptsize (2) Ekers et al. (1989);}\\
\multicolumn{8}{l}{\scriptsize (3) Jones \& Mc Adam (1992).}\\
\end{tabular}
\end{center}
\end{table*}

Concerning the integrated optical properties, 
galaxies associated with FRI radio morphologies seem intrinsically brighter
and larger than those hosting FRII sources (Owen \& Laing 1989).
In particular, according to Prestage \& Peacock (1988), FRIs are
associated with giant ellipticals similar to first ranked galaxies
and can be found in the cores of rich clusters, while
FRIIs are identified with galaxies of lower luminosity and are usually
found in poorer
clusters.
At $z \sim0.5$ the situation changes and both FRI and FRII
sources seem to lie in similar rich environments (Hill \& Lilly 1991). 

Another important point concerns the very nature of galaxies hosting
radio sources. It is commonly assumed that these galaxies are giant
ellipticals or bulge-dominated systems.  However, their brightness
profiles often exhibit marked deviations from a pure de Vaucouleurs
($r^{1/4}$) law, suggesting the presence of additional components in
both the nuclear and the external regions. These may be ascribed to unresolved nuclear sources,
small inner disks,
extended halos
or underlying disks.

The presence of nuclear point sources in the cores of radio galaxies was 
first investigated by Smith \& Heckman (1989b)
who reported that the light profiles of radio galaxies are consistent 
with an $r^{1/4}$ law with, in some cases (18 \%), an additional
point source at the nucleus.

The study of the galaxies hosting radio sources
and of the relation between optical and radio properties of galaxies
is an important tool for understanding
the nature of radio galaxies. In order to elucidate 
some of the above mentioned points we have therefore
 undertaken an optical study of a large sample (79 sources) of low-redshift
 (z~$\leq$~0.12) radio galaxies.
At this redshift limit we are able to investigate in detail galaxy 
properties using ground-based telescopes data.
Our sample includes both FRI and FRII with radio
 power in the range $\sim1.8\times10^{23}-3.1\times10^{26}$ Watt/Hz at 2700 MHz.

We used R band imaging to derive morphological and photometric
properties of the radio galaxies in the sample.
These include structural (ellipticity, isophote twisting, Fourier coefficients, etc.)
and  integrated observed properties (isophotal and total magnitude,
effective radius, etc.).
In particular we investigated the shape of the luminosity profile
and determined the contribution of the (unresolved) nuclear source.

Details on the observations are given in:\\
1) Fasano et al. 1996 (Paper I), where we present the 
results for a subset of 29 radio sources of the sample for which imaging in
both R and B band is available.\\
2) Govoni et al. 1999 (Paper II), where we
present the observational results from R band imaging of the remaining galaxies
(50 objects) in the sample.

The plan of this paper is the following:
in Sect. 2 we describe the optical data and the radio morphologies
of the galaxies.
In Sect. 3 we concentrate on the surface brightness profiles and in
 particular on the decomposition of the luminosity
 profiles.
In Sect. 4 we describe the global properties of the galaxies
(absolute magnitudes, effective radius, $\mu _e-R_e$ relation, and colors).
In Sect. 5 we discuss the structural and morphological properties of the galaxies (ellipticity and isophote shape).
In Sect. 6 we describe some signatures of interactions between galaxies
(displacement of isophote and twisting).
Finally, in Sect. 7 we summarize the results and draw the main
conclusions of the work.

Throughout this paper we assume $H_{0}$=50 km sec$^{-1}$ Mpc$^{-1}$
and $q_{0}$=0.  

\section{Optical and radio data}
We have collected images in the Cousins R band for 79 radio galaxies at low redshift, belonging to a sample of 95 objects.

This sample is composed of radio
galaxies extracted 
from two complete surveys of radio 
sources: 
The first one is the 2.7-GHz 2-Jy all-sky survey of radio sources
 with S(2.7GHz)$>$2Jy by Wall \& Peacock (1985, hereafter WP), and the second one is 
the survey from Ekers et al.
(1989, hereafter EK).
The selection criteria of our sample are described in Paper I.

Detailed surface photometry and quantitative morphological analysis
were performed 
using the AIAP package (Fasano 1990). Due to its high degree 
of interactivity, the AIAP software turns out to be particularly suitable
for analyzing the morphology of galaxies embedded in high density
regions, such as rich cluster or compact galaxy groups.
The observational details and data reduction are described in Paper I and
Paper II.

For each object we have also determined the shape of the point spread 
function (PSF) combining the radial brightness
profiles of several stars of different magnitude in each CCD
frame.  This allowed us to properly determine the PSF both in the center and
in the wings.

For each object we report in Table 1
the object name, the redshift, and the parent sample
(column 1, 2, and 3), together with
 radio data information (column 4, 5, 6, 7 and 8).

Concerning the radio data, most of the radio images of the objects
belonging to the WP sample were obtained by Morganti et al. (1993)
with the Very Large Array (VLA) and the Australia Telescope Compact
Array (ATCA), while VLA radio images are available for most of the
objects belonging to the EK sample.

Sources are classified according to the Fanaroff and Riley (FR) scheme
(Fanaroff \& Riley, 1974).  Simply stated, the FRI sources are
dominated by emission from the compact core and jets. The radio lobes
in these sources are generally diffuse, and fade with the distance
from the central source. The FRII sources have the highest surface
brightness in the hotspots at the outer extremes of the lobes.

The radio morphology classification, given in column 4 of Table 1 (FRI
and FRII indicated as I and II respectively), is mainly taken on the
basis of the published radio images.  Obviously, to allow a reliable
classification of these radio sources according to the FR scheme,
require good radio images (not always available).  Objects with
transition properties or with an unclear classification are marked as
I/II.  We indicate with U one unresolved source.  Finally, in the few 
case of objects with no available radio images we report the likely
class (labeled with a question mark) based on the power-morphology
relationship.  For these cases we classified as FRI all
radio sources with luminosity less than $2\times10^{25}$ Watt/Hz at
178 MHz.

Columns 6, 7 and 8 contain the radio fluxes at 408 MHz, at 2.7 GHz,
and the core radio flux at 4.8 GHz respectively.  In some cases only
upper limits of the core flux are available.  For the sources
belonging to the list of EK we have taken the radio flux from this
reference, while for the remaining objects the radio fluxes were taken
from WP at 2.7 GHz, from the PKS catalogue at 408 MHz, and from
Morganti et al. (1993) for the radio core at 4.8 GHz.

The average radio power at 2700 MHz turns out to be $1.1\times10^{25}$ Watt/Hz.

\section{Surface brightness profiles}

\subsection {Decomposition of the luminosity profiles}
To the first order, radial brightness profiles $I(r)$ of elliptical
galaxies are well described by the de Vaucouleurs law (1948)
$\mu\propto r^{1/4}$:
\begin{displaymath}
I(r)=I_0exp\{-7.67[(r/r_e)^{1/4}-1]\},
\end{displaymath} 
where $I_0$ is the surface brightness at $r=0$ and $r_e$ is the
effective radius.
In the case of radio galaxies, however,  
deviations from this law are rather common, in particular when looking
at the core or at the outer faint regions (Smith \& Heckman 1989b;
Colina \& de Juan 1995).
These deviations may be described either using extra components (e.g. inner
 disks, exponential law) or adopting a more general law (e.g. the Sersic (1968)
law $\mu \propto r^{1/n}$) to represent the profile.  

For the aims of this work we used a standard $r^{1/4}$ law (which remains a
good description for the main body of most ellipticals) plus, when required,
a nuclear component described by a PSF and an exponential law
to account for extra emission in the outer regions $\mu\propto r$:
\begin{displaymath}
 I(r)=I_dexp[-r/r_d],
\end{displaymath}
where $I_d$ and $r_d$ are the surface 
brightness at $r=0$ and the scale-length of the exponential
 component respectively.
This decomposition of the luminosity profile enables us to identify the
 presence of other components and to quantify deviations from 
the $r^{1/4}$ law.
Following this approach we fitted the observed brightness profiles of each
galaxy using
different strategies, from visual inspection
of the overall residuals, to best fit with $\chi^2$ minimization.
Errors associated with the profiles were computed using the prescriptions
given by Fasano \& Bonoli (1990).

In Fig. 1 we show our best decomposition of the brightness profiles
while in Table 2 we give all the relevant parameters of the fit.
Columns 2, 3 and 5 contain the magnitudes of point source, bulge and
exponential component ($m_{ps}$, $m_{r^{1/4}}$, $m_{exp}$),
respectively.  The effective radius of the bulge ($r_e$) and the
radius of the disk ($r_d$) in arcsec are reported in columns 4 and 6.
Column 7 gives the value of $\chi^2$ for the best fit.  We list the
different components of the profiles for each object in column 8,
while in column 9 we give additional notes.

For 19 galaxies the luminosity profile is fairly well fitted by a 
simple de Vaucouleurs law (R14).  For 13 galaxies we were not able to 
represent the external part of the luminosity profile with a 
$r^{1/4}$ law only, because of the presence of large excesses over 
the extrapolation of the $r^{1/4}$ law.  In these cases we found an 
acceptable fit by adding an exponential component (EXP) to the 
$r^{1/4}$ one.  For 28 sources the luminosity profile shows an excess 
over the $r^{1/4}$ law in the central unresolved region of the 
galaxy.  This excess can be satisfactory described by an additional 
point source (PS, modeled by the PSF).  On the contrary, 11 galaxies 
show an evident flattening (ABS) with respect to the $r^{1/4}$ law in 
the inner part of the luminosity profile.  In 4 objects we found both 
a point source in the nucleus and an external exponential component, 
while in other 4 sources there is both a deficit of light in the 
central region and an exponential component.  We summarize in Table 3 
the statistics of the components for the observed objects.  We found 
that the frequency of the different components in the brightness 
profiles of radio galaxies is not dependent on the radio morphology. 

\begin{figure*}
\resizebox{18.3cm}{22cm}{\includegraphics {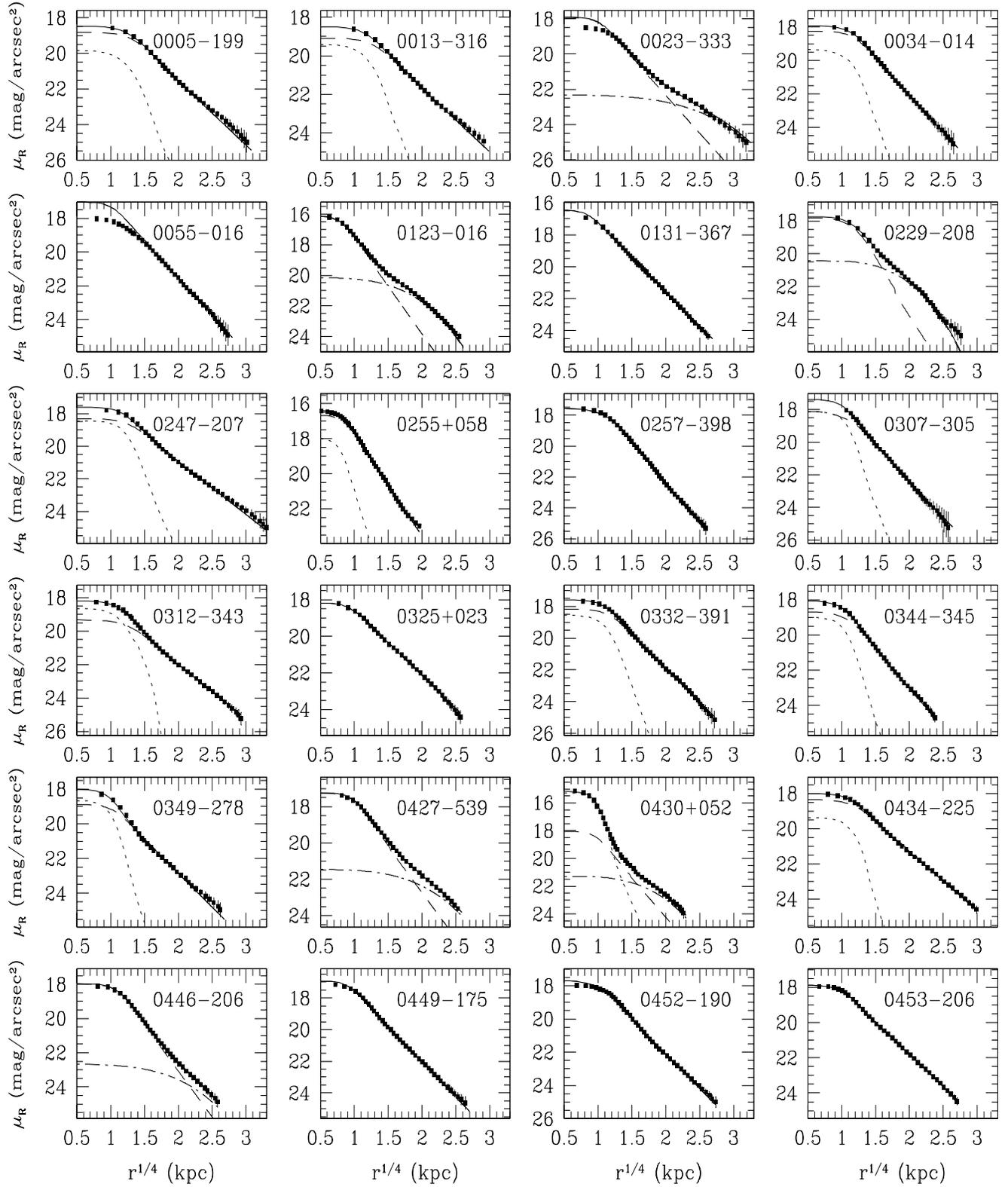}}
\hfill
\caption{Decomposition of the luminosity profile (filled dots in the figures)
 of the galaxies. These profiles are fitted with different components: point source (short dashed line), de Vaucouleurs law (long dashed line), exponential component (short dashed - long dashed line).}
\label{PSFig1_1}
\end{figure*}
\begin{figure*}
\resizebox{18.3cm}{22cm}{\includegraphics {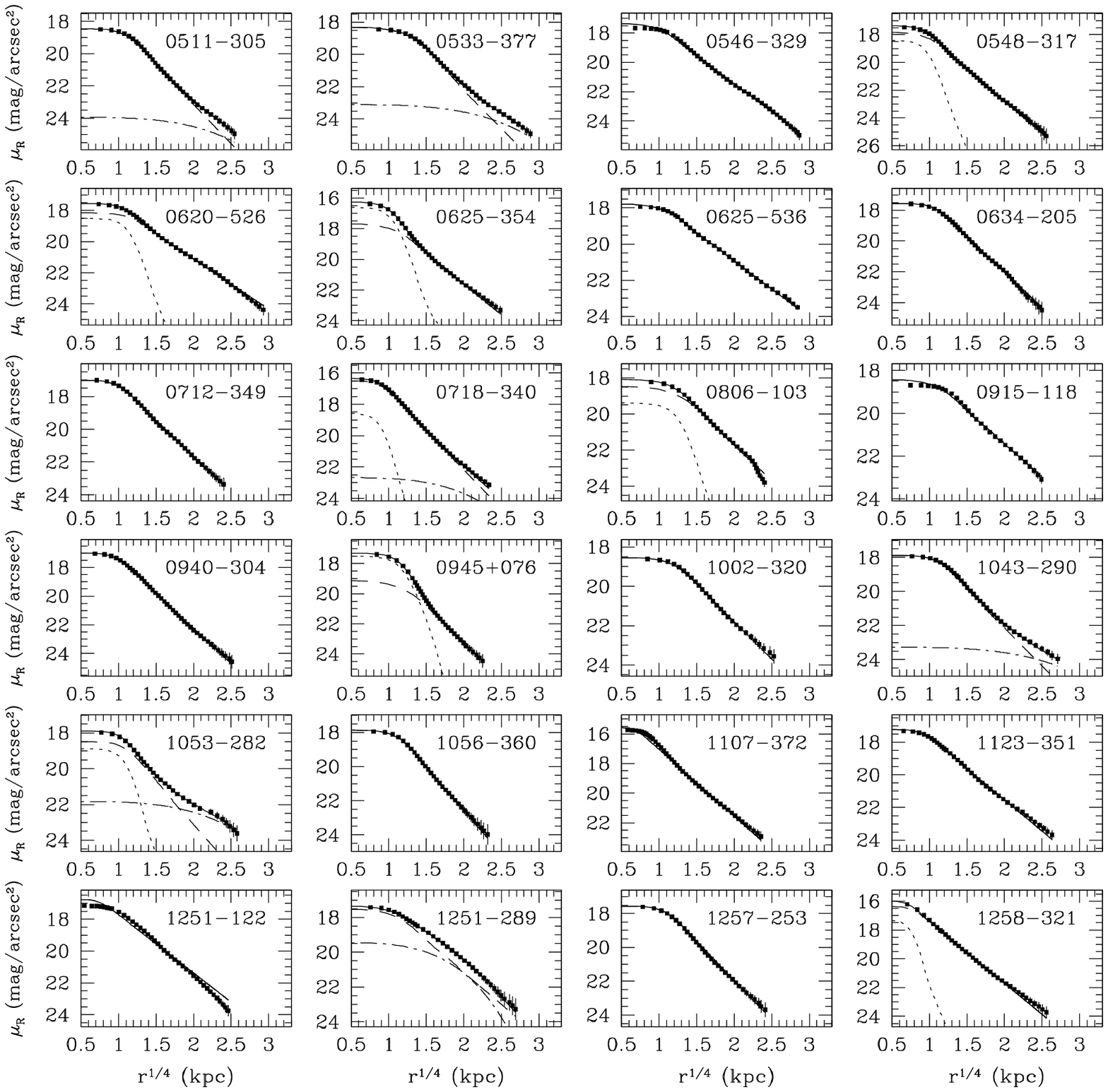}}
\hfill
{\bf Fig.~1.}~~Continued.
\end{figure*}
\begin{figure*}
\resizebox{18.3cm}{22cm}{\includegraphics {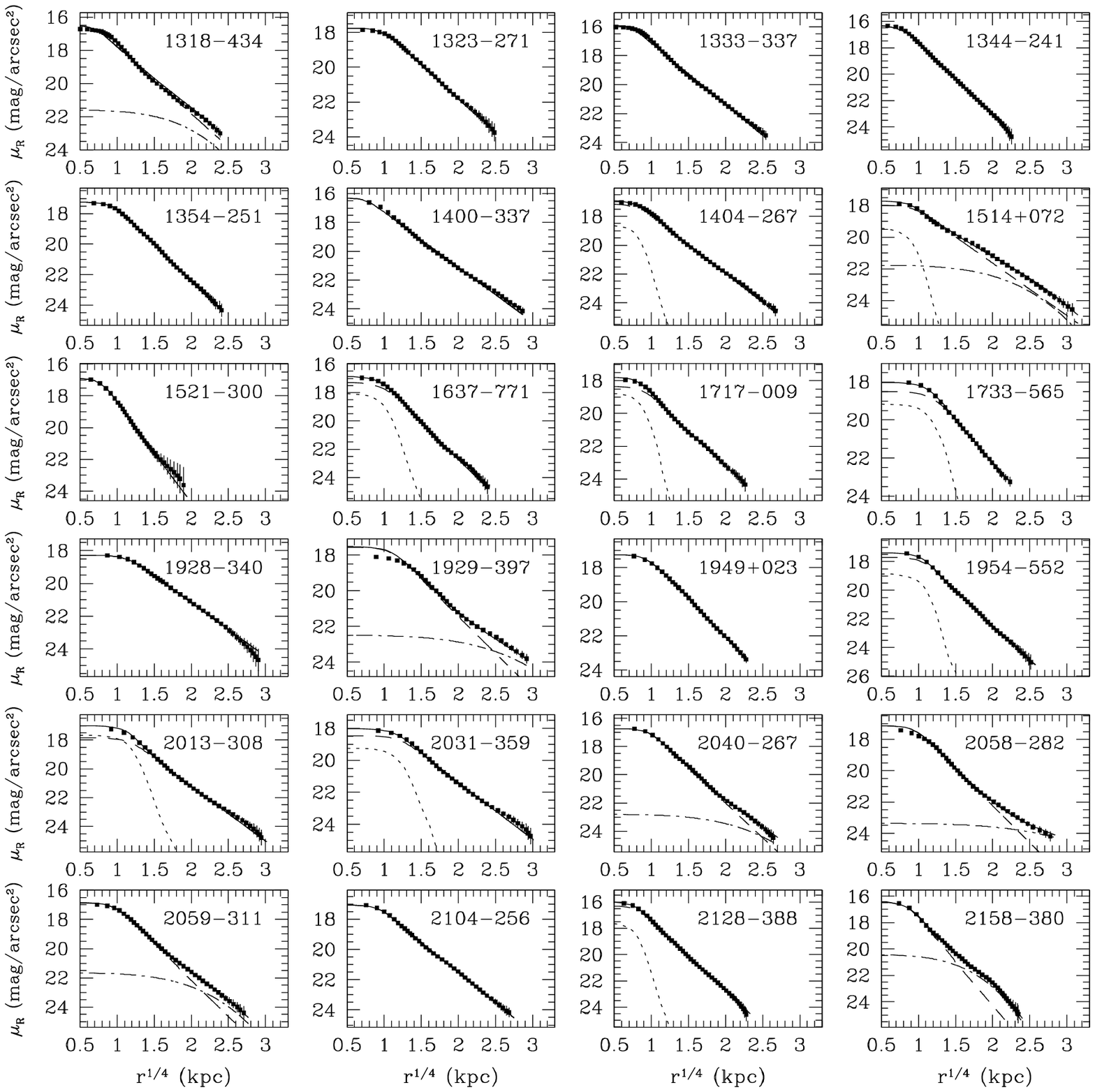}}
\hfill
{\bf Fig.~1.}~~Continued.
\end{figure*}
%
\begin{figure*}
\resizebox{18.3cm}{8.5cm}{\includegraphics {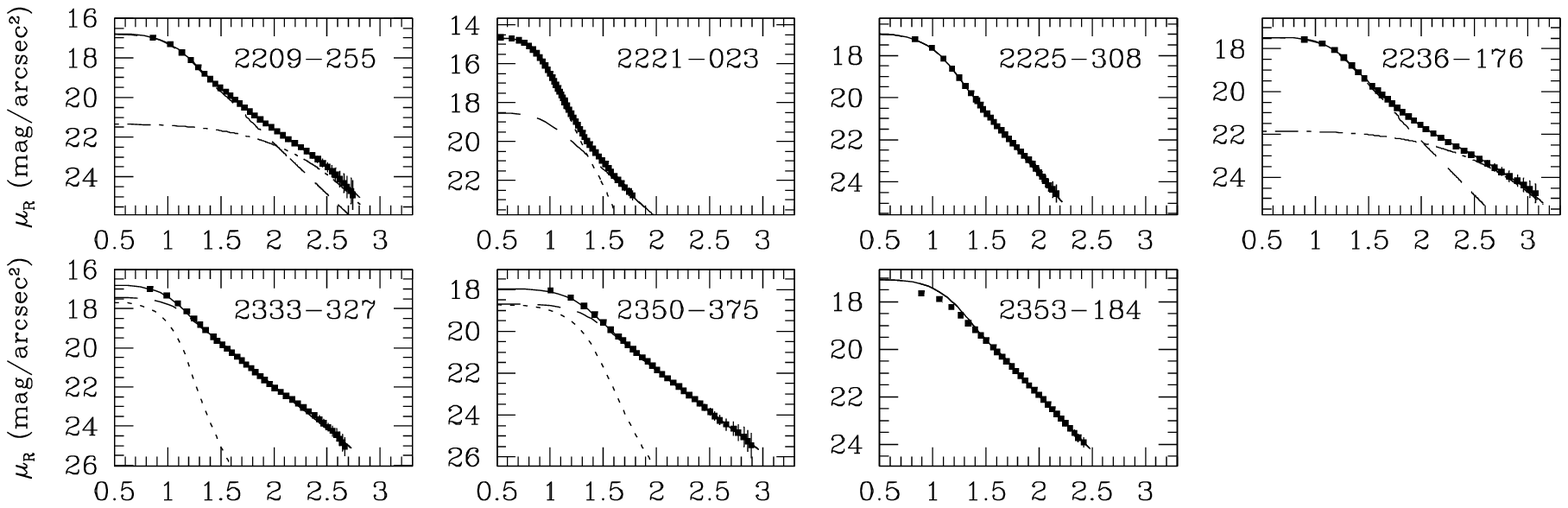}}
\hfill
{\bf Fig.~1.}~~Continued.
\end{figure*}

\subsection{Nuclear sources}

One interesting result of the present analysis of the luminosity
profiles is that $\sim$ 40\% of objects in our sample suggests the
presence of a nuclear point source.  In particular in some cases
(2221-023, 0430+052, 0625-354, 0945+076) this component is very
prominent.
 
Most of the investigations of radio galaxies in the optical band have
not tried to quantitatively determine the contributions of the nuclear
emission.  Only Smith \& Heckman (1989b) in their study of 72 powerful
radio galaxies have searched for point sources in the nuclei through
image modeling.  They used the results mainly to derive correction
factors for the galaxy magnitudes.  Excluding few peculiar sources
(e.g. 3C371 which is a BL Lac object), they found point sources for 13
cases with magnitude $M_V$ ranging from about -18 to -21 ($H_0=50$)
and contributing to the total flux of the object by $\sim 10\%$ or
less.  This is consistent with the distribution of the nucleus/galaxy
luminosity ratio we have found in our sample.  This distribution is
shown in Fig. 2 and peaks at Log[Nucl/Host]$\simeq -1.4$.  Therefore
in our sample of radio galaxies the average contribution of the
nuclear source to the galaxy luminosity is $\sim$ 5\%.  Assuming that
most of radio galaxies have a nuclear point source and that we are
detecting only the brightest nuclei, the average ratio nucleus/host
optical luminosity could be of the order of few percent.

%
\begin{table*}
\begin{center}
\begin{tabular}{ccccccccc}
\multicolumn{9}{c}{{\bf Table 2.} Parameters of the radial profile fit}\\
\multicolumn{9}{c}{}\\
\hline\\
\multicolumn{1}{c}{Name} & m$_{ps}$ & m$_{r^{1/4}}$ & r$_e$('') & m$_{exp}$ & r$_d$('') & $\chi^2$ & Components$^a$& Notes\\
\medskip
      (1)     & (2)	& (3)     & (4)         & (5)    & (6)          & (7)    &(8)           &(9) \\   
\hline\\
$    0005-199$&    18.84&    15.12&       8.1   &     ...&    ...     &    1.4 & R14,PS      &\\      
$    0013-316$&    18.38&    14.97&     12.9   &     ...&    ...      &    0.3 & R14,PS      &\\      
$    0023-333$&    ...  &    13.92&       9.8   &   13.44&    30.0    &   14.3 & R14,ABS,EXP  &halo\\  
$    0034-014$&    18.11&    14.68&      6.3   &     ...&    ...      &    0.2 & R14,PS      &\\      
$    0055-016$&    ...  &    13.07&      9.7   &     ...&    ...      &    ... & R14,ABS      &\\      
$    0123-016$&    ...  &    12.52&      6.0   &   11.75&   20.7      &    0.1 & R14,EXP      &disk\\  
$    0131-367$&    ...  &    12.15&     16.7   &     ...&    ...      &    1.5 & R14,ABS      &\\      
$    0229-208$&    ...  &    15.89&      1.1   &   15.24&    4.1      &    0.3 & R14,EXP      &possible disk\\  
$    0247-207$&    17.02&    13.62&     17.1   &     ...&    ...      &    0.9 & R14,PS      &\\      
$    0255+058$&    17.35&    13.57&      8.4   &     ...&    ...      &    0.7 & R14,PS      &\\      
$    0257-398$&    ...  &    14.54&      3.7   &     ...&    ...      &    1.0 & R14          &\\      
$    0307-305$&    17.26&    14.76&      6.9   &     ...&    ...      &    0.3 & R14,PS      &\\      
$    0312-343$&    16.88&    14.36&     22.5   &     ...&    ...      &    0.5 & R14,PS      &\\      
$    0325+023$&    ...  &    13.22&     32.2   &     ...&    ...      &    0.9 & R14          &\\      
$    0332-391$&    17.09&    14.17&      8.6   &     ...&    ...      &    0.2 & R14,PS      &\\      
$    0344-345$&    17.73&    14.98&      8.2   &     ...&    ...      &    0.1 & R14,PS      &\\      
$    0349-278$&    18.33&    15.23&     10.9   &     ...&    ...      &    2.6 & R14,PS      &\\      
$    0427-539$&    ...  &    13.71&      6.7   &   13.94&   17.1      &    0.2 & R14,EXP      &possible disk\\  
$    0430+052$&    14.31&    15.05&      5.0   &   14.92&   10.2      &    0.3 & R14,PS,EXP  &disk\\  
$    0434-225$&    17.91&    13.85&     18.9   &     ...&    ...      &    0.7 & R14,PS      &\\      
$    0446-206$&    ...  &    15.41&      2.7   &   16.51&    9.1      &    0.4 & R14,EXP     &halo\\      
$    0449-175$&    ...  &    12.77&     13.6   &     ...&    ...      &    1.5 & R14,ABS      &\\       
$    0452-190$&    ...  &    13.41&     14.0   &     ...&    ...      &    2.7 & R14,ABS      &\\      
$    0453-206$&    ...  &     12.9&     26.5   &     ...&    ...      &    0.2 & R14          &\\      
$    0511-305$&    ...  &    15.07&      5.2   &   16.76&   17.3      &    1.7 & R14,EXP    &halo\\     
$    0533-377$&    ...  &    15.27&      3.5   &   16.32&   13.3      &    0.5 & R14,EXP    &halo\\      
$    0546-329$&    ...  &    12.54&     21.7   &     ...&    ...      &    2.2 & R14,ABS     &\\     
$    0548-317$&    17.08&    13.67&     11.0   &     ...&    ...      &    0.2 & R14,PS      &\\     
$    0620-526$ &    16.88 & 12.94 &     30.2   &     ...&    ...      &    0.3 & R14,PS    &\\               
$    0625-354$ &    15.77 & 13.95 &     12.2   &     ...&    ...      &    0.4 & R14,PS      &\\               
$    0625-536$ &    ...   & 12.97 &     33.1   &     ...&    ...      &    0.2 & R14,ABS     &\\               
$    0634-205$ &    ...   & 14.82 &      6.7   &     ...&    ...      &    1.5 & R14         &\\               
$    0712-349$ &    ...   & 13.98 &     10.1   &     ...&    ...      &    0.7 & R14         &\\            
$    0718-340$ &    18.48 & 13.53 &      9.5   &   15.99&   20.0      &    0.4 & R14,PS,EXP &halo\\               
$    0806-103$ &    18.81 & 15.8  &      6.8   &     ...&    ...      &    0.9 & R14,PS      &\\               
$    0915-118$ &    ...   & 14.48 &     33.9   &     ...&    ...      &    4.0 & R14,ABS     &\\               
$    0940-304$ &    ...   & 13.69 &      7.7   &     ...&    ...      &    1.0 & R14         &\\               
$    0945+076$ &     16.6 & 16.39 &      4.6   &     ...&    ...      &    0.3 & R14,PS      &\\               
$    1002-320$ &    ...   & 15.3  &      9.5   &     ...&    ...      &    0.3 & R14         &\\               
$    1043-290$ &    ...   & 14.38 &      7.0   &   15.56&   30.0      &    0.6 & R14,EXP     &halo\\               
$    1053-282$ &    17.97 & 15.36 &      5.4   &   14.99&   15.9      &    0.2 & R14,PS,EXP &possible disk\\ 
$    1056-360$ &    ...   & 15.16 &      3.9   &     ...&    ...      &    0.1 & R14         &\\               
$    1107-372$ &    ...   & 10.16 &      40.4   &     ...&    ...     &    8.1 & R14,ABS     &\\               
$    1123-351$ &    ...   & 12.61 &      23.8   &     ...&    ...     &    0.6 & R14         &small~halo\\      
$    1251-122$ &    ...   & 10.83 &      61.9   &     ...&    ...     &    9.4 & R14,ABS     &\\               
$    1251-289$ &    ...   & 13.47 &      16.2   &   13.91&   5.9      &    0.1 & R14,EXP     &halo \\
$    1257-253$ &    ...   & 14.55 &       6.0   &     ...&    ...     &    1.5 & R14         &\\               
$    1258-321$ &    16.65 & 11.35 &     28.3   &     ...&    ...      &    0.5 & R14,PS     &small~halo\\       
$    1318-434$ &    ...   & 10.79 &       58.0   &    12.2&  45.0     &    5.3 & R14,ABS,EXP  &halo\\               
$    1323-271$ &    ...   & 13.62 &      17.3   &     ...&    ...     &    0.7 & R14         &\\               
$    1333-337$ &    ...   & 10.38 &      39.5   &     ...&    ...     &    2.5 & R14,ABS     &\\               
$    1344-241$ &    ...   & 12.71 &      9.5   &     ...&    ...      &    0.4 & R14         &\\               
$    1354-251$ &    ...   & 13.71 &       8.5   &     ...&    ...     &    0.3 & R14         &\\        
\hline\\					     		  	  
\end{tabular}				
\end{center}								
\end{table*}

\begin{table*}
\begin{center}
\begin{tabular}{ccccccccc}
\multicolumn{9}{c}{{\bf Table 2 (continued).} Parameters of the radial profile fit}\\
\multicolumn{9}{c}{}\\
\hline\\
\multicolumn{1}{c}{Name} & m$_{ps}$ & m$_{r^{1/4}}$ & r$_e$('') & m$_{exp}$ & r$_d$('') & $\chi^2$ & Components$^a$& Notes\\
\medskip
                 (1)     & (2)	& (3) & (4)& (5) & (6)& (7) &(8)&(9)    \\
\hline\\
$    1400-337$&    ...    & 10.25 &      56.8    &    ...&    ...     &    3.9 & R14         &\\        $    1404-267$&    17.85&    12.09&    29.0    &         ...&     ...    &        0.7  & R14,PS     &\\     
$    1514+072$&    18.73&    12.51&     39.2    &       13.12&     25.0  &        3.2  & R14,PS,EXP &possible disk\\ 
$    1521-300$&    ...  &    14.06&     5.4    &         ...&     ...    &        0.6  & R14        &small halo\\
$    1637-771$&    17.18&    14.12&     6.0    &         ...&     ...    &        0.6  & R14,PS     &\\     
$    1717-009$&    18.42&    14.53&    16.4    &         ...&     ...    &        0.6  & R14,PS     &\\     
$    1733-565$&    18.74&    16.09&     5.4    &         ...&     ...    &        0.2  & R14,PS     &\\     
$    1928-340$&    ...  &    14.58&    15.1    &         ...&     ...    &        0.1  & R14         &\\     
$    1929-397$&    ...  &    14.15&     8.0    &       14.94&     22.0     &     15.3  & R14,ABS,EXP &halo\\     
$    1949+023$&    ...  &    14.86&     6.9    &         ...&     ...    &        0.1  & R14         &\\     
$    1954-552$&    17.95&    14.43&     5.6    &         ...&     ...    &        0.1  & R14,PS     &\\     
$    2013-308$&    17.01&    14.37&     9.1    &         ...&     ...    &        1.9  & R14,PS     &\\     
$    2031-359$&    18.11&    14.37&    12.2    &         ...&     ...    &        1.2  & R14,PS     &\\ 
$    2040-267$&    ...  &    13.27&     6.9    &       14.98&   22.3    &         0.7  & R14,EXP     &halo\\     
$    2058-282$&    ...  &    13.53&     8.6    &       14.52&   68.9    &         5.2  & R14,ABS,EXP &halo\\     
$    2059-311$&    ...  &    13.37&     7.1    &        14.2&   16.2    &         0.2  & R14,EXP    &possible disk\\
$    2104-256$&    ...  &    12.84&    17.1    &         ...&     ...    &        0.6  & R14        &\\     
$    2128-388$&    16.66&    12.11&    12.2    &         ...&     ...    &        0.2  & R14,PS    &\\     
$    2158-380$&    ...  &    14.13&     2.7    &        14.3&    7.1    &         0.5  & R14,EXP    &disk\\ 
$    2209-255$&    ...  &    14.26&      3.8    &        14.8&    9.1    &        0.1  & R14,EXP    &disk\\ 
$    2221-023$&    15.08&    16.17&     5.3    &         ...&     ...    &        0.2  & R14,PS    &\\     
$    2225-308$&    ...  &    15.03&      2.4    &         ...&     ...    &       0.4  & R14        &\\     
$    2236-176$&    ...  &    14.55&     3.6    &       14.34&    15.1   &         0.1  & R14,EXP    &possible disk\\ 
$    2333-327$&    17.08&    13.72&     9.1    &         ...&     ...    &        0.6  & R14,PS    &\\     
$    2350-375$&    17.83&    15.34&     5.9    &         ...&     ...    &        0.3  & R14,PS    &\\     
$    2353-184$&    ...  &    14.46&     4.8    &         ...&     ...    &        8.4  & R14,ABS    &\\     
\hline\\
\multicolumn{9}{l}{\scriptsize $^a$ Decomposition of the brightness profiles:}\\
\multicolumn{9}{l}{\scriptsize R14 = de Vaucouleurs law; PS = unresolved component in the core of the source;}\\ 
\multicolumn{9}{l}{\scriptsize ABS = lack of light with respect to the de Vaucouleurs law in the center;}\\
\multicolumn{9}{l}{\scriptsize EXP = exponential component in the external region.}\\
\end{tabular}								
\end{center}
\end{table*}

Recent studies of radio galaxies using Hubble Space Telescope (HST)
observations have confirmed the presence of bright optical nuclear
unresolved components
in galaxies hosting radio sources (Capetti \& Celotti 1999; McLure et
al. 1999; Chiaberge et al. 1999).

In BL Lac objects generally the nuclear source has a luminosity
similar to that of the galaxy (Falomo et al. 1999), while in quasars
the nucleus/host luminosity ratio is of the order of $\sim$10.

The distribution of the absolute magnitudes of the point sources is
shown in Fig. 3a ($\langle M_{PS}\rangle=-20.4\pm1.3$; errors are
given as the r.m.s throughout the paper), while Fig. 3b shows that the
magnitude of the point source is not correlated with that of the host
galaxy.

\begin{table*}
\begin{center}
\begin{tabular}{ccccccc}
\multicolumn{7}{c}{{\bf Table 3.} Statistic of galaxy components}\\
\multicolumn{7}{c}{}\\
\hline\\
\multicolumn{1}{c}{}&\multicolumn{1}{c}{R14}
&\multicolumn{1}{c}{R14,PS} &\multicolumn{1}{c}{R14,PS,EXP} &\multicolumn{1}{c}{R14,ABS}&\multicolumn{1}{c}{R14,ABS,EXP} &\multicolumn{1}{c}{R14,EXP}\\ 
\multicolumn{7}{c}{}\\
\hline\\
  All                 &       &       &       &         &         &        \\
N. of objects         & 19    & 28    & 4     & 11      & 4       & 13     \\
\% of Objects         & 24\%  & 35.5\%& 5\%   & 14\%    & 5\%     & 16.5\% \\
                      &       &       &       &         &         &        \\ 
  FRI                 &       &       &       &         &         &        \\
N. of objects         & 10    & 16    & 3     &  8      & 2       & 8      \\
\% of objects         & 21.5\%& 34\%  & 6\%   &  17\%   & 4.5\%   & 17\%   \\
                      &       &       &       &         &         &        \\
  FRII                &       &       &       &         &         &        \\
N. of objects         &  3    & 10    & 0     &  2      & 1       &  3     \\
\% of objects         & 16\%  & 52.5\%& 0\%   &  10.5\% & 5\%     & 16\%   \\
\hline\\

\multicolumn{7}{l}{\scriptsize R14 : de Vaucouleurs law; PS : an unresolved component in the core of the source;}\\ 
\multicolumn{7}{l}{\scriptsize ABS : a lack of light with respect to the de Vaucouleurs law in the center;}\\
\multicolumn{7}{l}{\scriptsize EXP : an exponential component in the external region.}\\
\end{tabular}								
\end{center}
\end{table*}

We derived an  upper limit for the point source contribution in the
galaxies where no point source is detected.
This was done by adding a point source component at various levels and seeing
when the contribution would became observable (see Table 5, column 8) . 

From Fig. 4, the radio core power appears weakly correlated with
the luminosity of the point source.
A linear correlation of the nuclear optical luminosity with the radio core
was also recently reported for a sample of 33 FRI
radio galaxies from 3CR catalogue observed with HST (Chiaberge et al. 1999).

Giovannini et al. (1988) found a relation between total radio power
and core radio power ($LogP_{core}=11.01+0.47LogP_{tot}$) in a sample
of 187 radio galaxies selected at low frequency.  In Fig. 5 we plot
the core radio power at 4800 MHz versus the total radio power at 408
MHz; our objects are divided according to the PS absolute magnitude.

The radio core power of our radio galaxies are systematically brighter
with respect to this empirical relationship.  This is probably due to
the fact that our sample is biased toward high radio core power due to
the high frequency (2700 MHz) at which the sample was selected.  The
scatter in the relation is attributed to the different orientation of
the sources with respect to the line of sight (Giovannini et
al. 1988).  Namely, sources pointing toward the observer have their
radio core powers enhanced by Doppler boosting.  Consistent with this
view we note that the objects with more luminous optical point sources are
located towards higher radio core power with respect to the objects
for which no point source was detected (see Fig. 5).

\begin{figure}
\resizebox{\hsize}{!}{\includegraphics {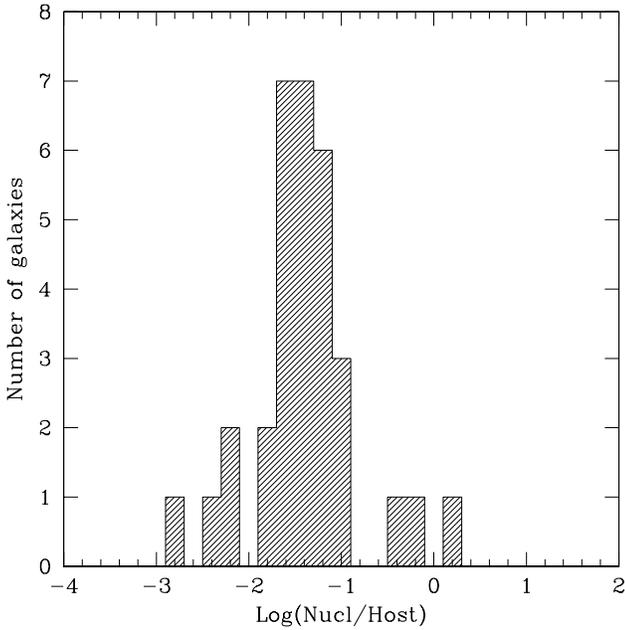}}
\hfill
\caption{Distribution of the nucleus/galaxy luminosity ratio($Log[Nuc/Host]$)
in our sample.}
\label{PSistPSF}
\end{figure}
\begin{figure}
\resizebox{\hsize}{!}{\includegraphics {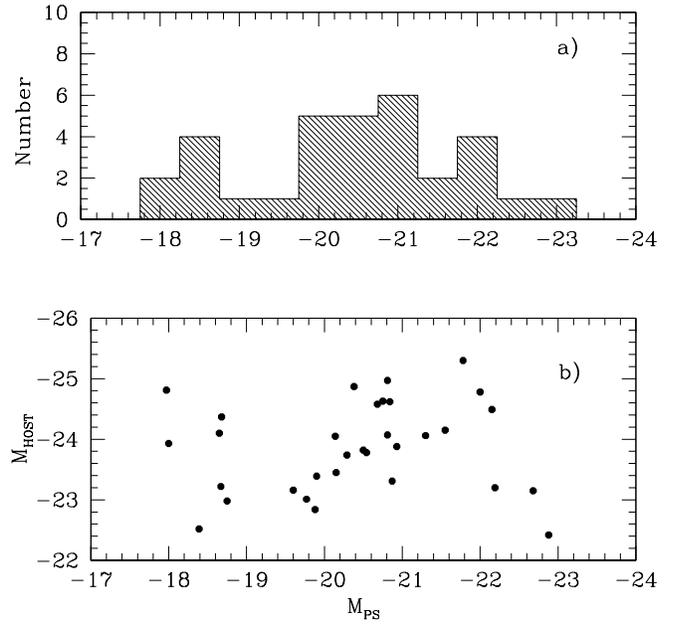}}
\hfill
\caption{a) Histogram of the absolute magnitude of the point source;
 b) Absolute magnitudes of the host galaxies ($M_{HOST}$) plotted against
the absolute magnitudes of the point sources $M_{PS}$.}
\label{PSpsf}
\end{figure}
\begin{figure}
\resizebox{\hsize}{!}{\includegraphics {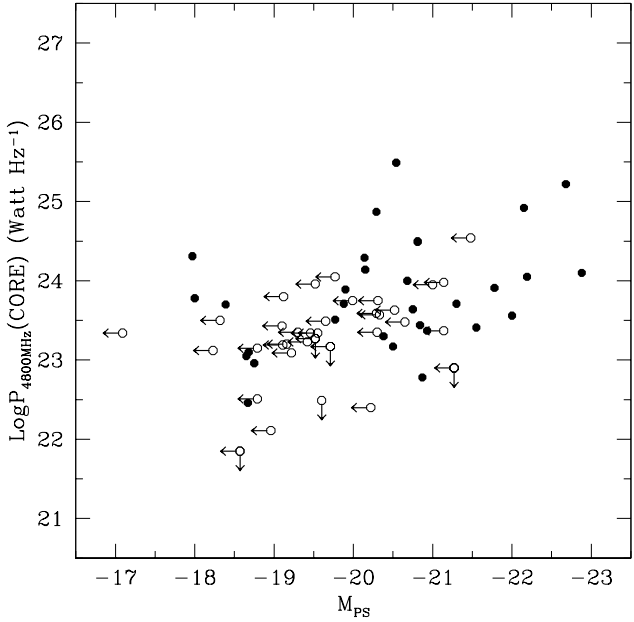}}
\hfill
\caption{Relation between core radio power at 4800 MHz and absolute optical
 magnitude of the point source. Filled circles represent objects with
 detected optical point sources. Open circles indicate upper
limits for the point source contribution or for the core radio power
 (see text for details).
}
\label{PSradPSF}
\end{figure}
\begin{figure}
\resizebox{\hsize}{!}{\includegraphics {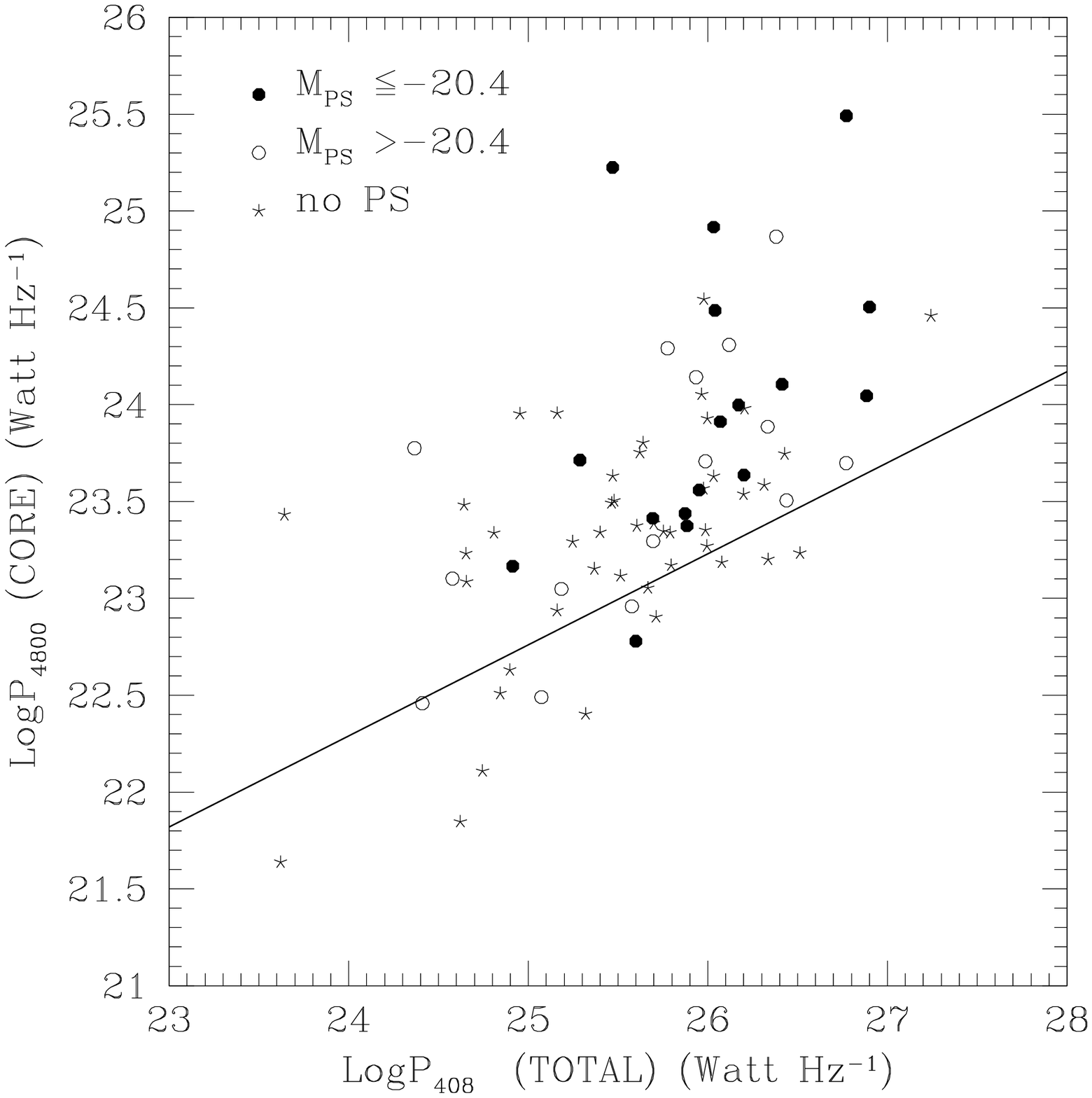}}
\hfill
\caption{Relation between total radio power at 408 MHz and core radio power
at 4800 MHz. Objects are divided according to luminosity of PS: $M_{PS}<-20.4$ (filled circles),
$M_{PS}>-20.4$ (open circles), and undetected PS (stars).
The solid line
represents the empirical relation found by Giovannini et al. (1988).} 
\label{PSRadiocore}
\end{figure}

\subsection{Inner deficit of light} 

The luminosity profiles of 15 sources (5 cases are clearly evident in
Fig. 1) show in the nuclear part a lack of light with respect to the
de Vaucouleurs fit. This could be ascribed either to a substantial
deviation from isothermal distribution of stars or to significant
diffuse dust absorption.  Wise \& Silva (1997) examined the effects of
dust on the central optical properties of elliptical galaxies. They
found that smoothly distributed dust will significantly suppress the
surface brightness of the central region of ellipticals without being
clearly visible in direct images.  However this interpretation would
imply a clear signature in the color profiles, which is not observed.
On the other hand the presence of dust in the central regions of radio
galaxies is often detected from HST images  (Chiaberge et al. 1999; Kleijn
et al. 1999)  as dust lanes or dusty-disks. These features are,
however, in general confined in a region of about 1 kpc and are 
therefore undetectable in ground based observations. From our images
only a couple of galaxies show clear presence of dust features (see
Paper I and Paper II for details).

\subsection{External light excess}

A significant fraction of radio galaxies in our sample ($\sim25\%$)
exhibit an excess over the $r^{1/4}$ fit for radii $r\geq r_e$.
Similar light excesses have been reported by Colina \& de Juan (1995)
for most of the FRI radio galaxies in their sample, and ascribed to
the effects of galaxy interactions.  While this interpretation could
apply to some of our galaxies, we explore different interpretations
for the excess; namely a disk or halo component.

In the last column of Table 2 we speculate about the nature of the
outer light excess with respect to the $r^{1/4}$. The excess has been
modeled with an exponential component.  The nature of this component
depends on several factors.  Besides the very appearance of the
galaxy, one speculation usually relies on different "objective" clues
(see also Paper II): a) the central surface brightness of the
exponential component, which, in the case of disks, according to
Freeman (1970), should span the range 20--20.75 in the R band; b) the
$c_4$ profile, whose positive sign should indicate the presence of a
disk component; c) ellipticities and position angles of the isophotes, which, in
the ideal case of thin disks superimposed to ellipsoidal or biaxial
bulges, should exhibit increasing or constant ellipticity profiles,
respectively. On the basis of these criteria, we found 4  
galaxies in which the presence of an outer disk component turns out to
be very likely.  We also found 11 cases in which the light excesses can be
confidently ascribed to the presence of large outer halos.  The note
"small halo" in Table 2 has been introduced to denote 3 galaxies for
which no exponential component has been used in the fit, since the
amount of the light excess in the outer region is rather small.
Finally, there are 6 cases in Table 2 for which the profile is rather complex
and no good fit can be obtained using simple models.  In these cases we are not able to
decide about the nature of the light excess in the luminosity profiles. 

The fraction of galaxies in our sample showing a light excess in
the outer profiles is of 25\%, a percentage much smaller
than that found by Colina \& de Juan (1995) (73\%).  If the excess is
attributed to galaxy interactions our results would imply they are
significantly less frequent than previously reported and comparable to
normal ellipticals.

\section{Overall properties of host galaxies}

\subsection{Absolute magnitudes and effective radii.}
\begin{figure}
\resizebox{\hsize}{!}{\includegraphics {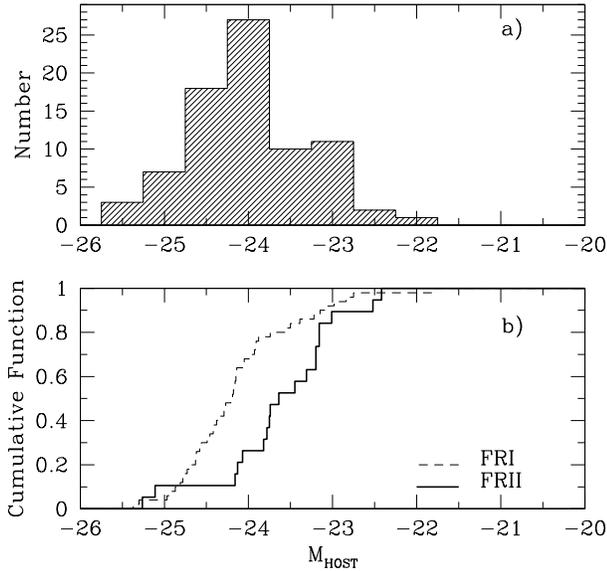}}
\hfill
\caption{a) Distribution of absolute total magnitude
of the host galaxies in R band. b) Cumulative distribution of the absolute
total magnitude of the host galaxies for FRI (dotted line) and FRII
 (solid line).}
\label{PSmag1}
\end{figure}
We investigate here the optical luminosity and the effective radius
of radio galaxies and
compare our results with previous findings from other samples of
radio galaxies as well as with normal elliptical galaxies.

In Table 5 we report the photometric and structural properties of the
observed objects.  Column 1 gives the name of the radio galaxy.
Column 4 contains the integrated magnitudes of the objects computed
within the isophote 24.5.  We computed the total magnitudes (column 5)
from extrapolation of the surface brightness profiles and we used
these values to derive the effective radii $R_e$ (semimajor axis of
the isophote enclosing half of the total galaxy light; column 2) and
the corresponding effective surface brightness $\mu_e$ (column 3). In
Table 5 we give the magnitudes and the surface brightness corrected
for galactic extinction.  The absolute R band magnitudes
$M_{HOST}(24.5)$ and $M_{HOST}(tot)$ reported in Table 5 (column 6 and
7) include K-correction (see Paper I-II). Moreover these magnitudes
have also been corrected for the contribution of the nuclear component
derived from the fit (see Table 5, column 8).

We found that the average absolute magnitude is $\langle 
M_{HOST}(tot)\rangle=-24.00\pm 0.69$.  The distribution of absolute 
magnitudes for the whole data-set is presented in Fig. 6a. 
 
We compare our finding on the absolute magnitudes of radio galaxies
with the results by Owen \& Laing (1989), Smith \& Heckman (1989b) and
Ledlow \& Owen (1995).  In order to perform a suitable comparison we
have transformed their data to our cosmology.  In addition since many
authors publish only isophotal magnitudes (at $\mu_R =24.5$ or $\mu_V
=25$) we applied an average correction of $-0.17$ mag to transform
isophotal magnitudes ($m_{24.5}$) in total magnitudes, an average
correction of $-0.12$ mag to transform isophotal magnitudes ($m_{25}$)
into total magnitudes (these values are taken from our average
difference between $m_{24.5}$ and $m_{tot}$ and between $m_{25}$ and
$m_{tot}$).  A summary of the comparison of our absolute magnitudes
with those from other samples of radio galaxies is given in Table 4,
which reports the samples (divided also between FRI and FRII), the
mean redshift of the samples, and in brackets, the number of objects
in each sample.  The agreement between our absolute magnitudes with
those from other sample is good.
\begin{figure}
\resizebox{\hsize}{!}{\includegraphics {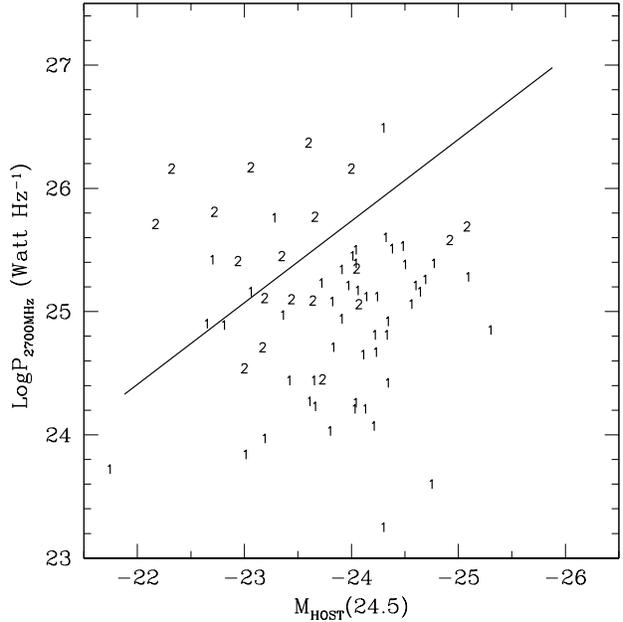}}
\hfill
\caption{The radio power at 2700 MHz vs absolute magnitude
of the host galaxy. FRI and FRII are indicated as 1 and 2 respectively.
 The line in the plot represents the segregation
between radio power and optical luminosity proposed by Owen \& Ledlow (1994).}
\label{PSradmag}
\end{figure}
Dividing the radio galaxies according to their radio morphology (FRI
and FRII; see previous section) we find that galaxies hosting FRI
sources are systematically brighter than those hosting FRII sources. A
systematic difference in luminosity between FRI and FRII galaxies was
previously suggested by Owen \& Laing (1989) from observations of 47
radio galaxies in the R-band.

Including only the objects with good morphological radio
classification we found: $\langle M_{HOST}(tot)\rangle _{FRI}=-24.13
\pm 0.69$ and $\langle M_{HOST}(tot)\rangle _{FRII}=-23.62 \pm 0.73$.
This difference is illustrated in Fig. 6b by comparison of the two
cumulative distributions.  The Kolmogorov-Smirnov two-samples test
yields in this case P$_{KS}$ = 0.001.

Owen \& Laing (1989) and Owen \& White (1991) have noted a segregation
of FRI and FRII radio galaxies in the radio power to host optical
luminosity plane, suggesting that segregation between FR I/II sources
was not only dependent on the radio power but also on the optical
luminosity of the host galaxy.  We report in Fig. 7 the total radio
luminosity at 2700 MHz versus $M_{HOST}(24.5)$ for our sample. The
line in the plot represents the dividing line proposed by Owen \&
Ledlow (1994) for FRI and FRII.

The scatter of our galaxies over this plane is very large, but the
figure suggests that the galaxies associated with different radio
morphology (FRI and FRII) fall generally in separate areas of the
diagram.  A result that, together with the classical radio luminosity
division, reflects the difference of average optical galaxy
luminosity.

\begin{table*}
\begin{center}
\begin{tabular}{ccccc}
\multicolumn{5}{c}{{\bf Table 4.} Total absolute R magnitudes of radio galaxies}\\
\multicolumn{5}{c}{}\\
\hline\\
\multicolumn{1}{c}{}&\multicolumn{1}{c}{This work}
&\multicolumn{1}{c}{Ledlow,Owen} &\multicolumn{1}{c}{Owen,Laing} &\multicolumn{1}{c}{Smith,Heckmann} \\
\multicolumn{5}{c}{}\\
\hline\\
All                      &$-24.00$ (79)&    -           & $-23.59$ (47)& $-23.72$ (68)    \\
$\langle z\rangle$       & 0.053       &    -           & 0.081        & 0.093           \\
                         &             &                &              &                 \\
FRI                      &$-24.13$ (50)&  $-23.83$ (265)& $-23.96$ (13)& $-23.87$ (25)    \\
$\langle z_{FRI}\rangle$ & 0.052       &   0.087        & 0.034        & 0.057           \\
                         &             &                &              &                \\ 
FRII                     &$-23.62$ (19) &    -          & $-23.30$ (26)& $-23.59$ (33)    \\
$\langle z_{FRII}\rangle$& 0.056       &    -           & 0.110        & 0.123           \\

\hline\\							
\end{tabular}
\end{center}
\end{table*}

The hosts of FRI radio sources also appear 
larger than FRII hosts. From our sample we find that the average effective
radius is 
$18.4\pm10.1$ kpc for FRIs and $12.9\pm9.3$ kpc for FRIIs
(see Fig. 8).
\begin{figure}
\resizebox{\hsize}{!}{\includegraphics {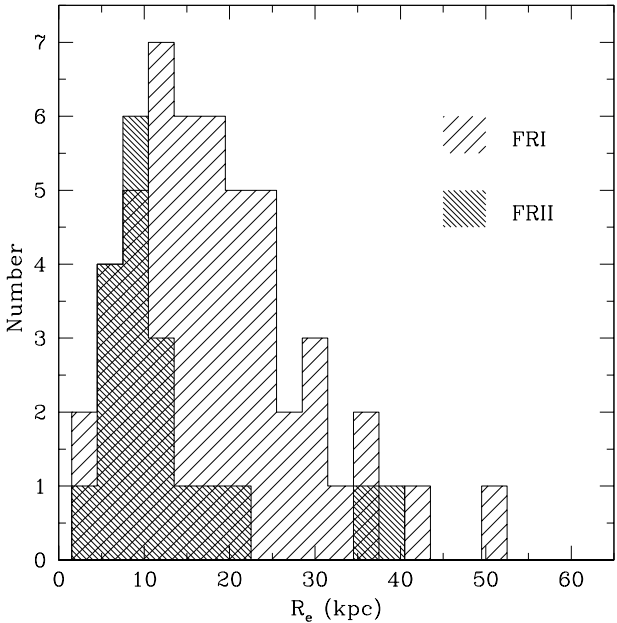}}
\hfill
\caption{Effective radius (in kpc) divided between FRI and FRII.}
\label{PSReff}
\end{figure}
This is likely to be simply a consequence of the relationship between
effective radius and luminosity of elliptical galaxies. For our 79
radio galaxies we find:

 $M_{HOST} =-20.77(\pm 0.28) -2.78(\pm 0.24)\times Log(R_e)$.\\
The same relation appears to apply for normal luminous
ellipticals (Romanishin 1986).

\subsection{$\mu _e-R_e$ relation}

Fig. 9a illustrates the relation $\mu_e - R_e$ (hereafter Kormendy
relation; Kormendy 1977) we obtained from our sample of radio
galaxies.  The effective surface brightnesses of the galaxies have
been corrected for galactic extinction and cosmological dimming
$10\times\log(1+z)$. Moreover, effective radii and surface
brightnesses of four galaxies with very bright nuclei (2221-02,
0625-35, 0430+05, 0945+07) have been properly corrected to take into
account the contribution of the point source.

The straight line best fitting the whole sample is given by the equation:

$\mu_e~({\rm All}) = 18.44(\pm 0.35) + 2.58(\pm 0.29)\times log(R_e)$,

which is consistent with the fit obtained in Paper I from our
sub-sample of 29 radio galaxies. The fits relative to the FRI and 
FRII sub-samples are given by:

$\mu_e(FRI) = 18.51(\pm 0.41) + 2.49(\pm 0.33)\times log(R_e)$,

$\mu_e(FRII) = 18.35(\pm 0.61) + 2.71(\pm 0.60)\times log(R_e)$.

They are represented in Fig. 9a by the full and dotted lines,
respectively. 

In conclusion, the coefficients of the different $\mu_e-R_e$ relations
do not differ significantly from one another.
\begin{figure}
\resizebox{\hsize}{!}{\includegraphics {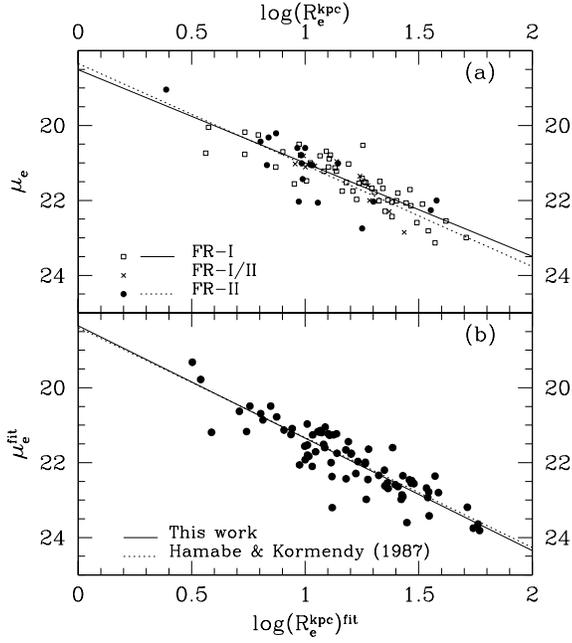}}
\hfill
\caption{a) $\mu_e - R_e$ relation.
The effective radii and surface brightnesses refer to the isophote enclosing half of the total galaxy light. The solid and dotted lines represent FRI and FRII radio source, respectively; b) $\mu_e - R_e$ relation obtained by fitting with
a de Vaucouleurs law the luminosity profile. The solid lines represent the relation relative to the global sample, the dotted lines represent the relation
by Hamabe \& Kormendy (1987) }
\label{PSmue_re}
\end{figure}

The effective radii and surface brightnesses in Fig. 9a refer to the
isophote enclosing half of the total galaxy light (column 2 of Table
5). We prefer to use these quantities rather than those obtained by
fitting the luminosity profiles of the galaxies with a de~Vaucouleurs
$r^{1/4}$ law because they are model independent.  However, since the
latter definitions of $R_e$ and $\mu_e$ are widely used in the
literature, in order to compare our results with those obtained from
other samples, we also computed them for our radio galaxies.  Fig. 9b
illustrates the $\mu_e - R_e^{fit}$ relation from our sample. In this
case the best fits to the data turn out to be:

$\mu_e~({\rm All}) = 18.35(\pm 0.24) + 3.00(\pm 0.19)\times log(R_e)^{fit}$,

$\mu_e(FRI) = 18.68(\pm 0.26) + 2.70(\pm 0.20)\times log(R_e)^{fit}$,

$\mu_e(FRII) = 17.66(\pm 0.69) + 3.71(\pm 0.65)\times log(R_e)^{fit}$.

Again the relation relative to the global sample (solid line in
 Fig. 9b) is fairly consistent with that obtained in Paper I for a
 much less sizeable sample.

We compared our results with those derived from other samples of radio
galaxies. After accounting for the different cosmology, we found a
good agreement with the results of Colina \& de~Juan (1995) for a
sample of FRI galaxies ($\mu_e=18.34+3.03\ log(R_e)$) and Smith \&
Heckman (1989b) for powerful radio galaxies ($\mu_e=18.22+3.03\
log(R_e)$; assuming a color correction $V-R=0.8$).  Instead a
significant difference is found with the $\mu_e - R_e$ relation given
by Ledlow \& Owen (1995) for a sample of FRI galaxies in Abell
clusters.  They report $\mu_e - R_e$ relationships for various
subsamples that systematically differ from ours for brighter zero
points and steeper slopes.  A possible explanation for this
discrepancy is that, at variance with our sample, the Ledlow \& Owen
galaxies live in rich environments.

It is well known that the Kormendy relation represents the projection
on the plane ($\mu_e - R_e$) of the fundamental plane of elliptical
galaxies (Djorgovski \& Davis 1987, Dressler et al. 1987). This
relation has been claimed to be closely related to the morphological
and dynamical structure of galaxies, as well as to their formation
processes.

It is therefore of interest to compare the Kormendy relation for radio
galaxies with that for normal ellipticals.  Comparing with the
classical $\mu_e - R_e$ relation given by Hamabe \& Kormendy (1987)
for radio-quiet ellipticals, ($\mu_e=18.39+2.94\ log(R_e)$ in our
system); we found the two relationships are indistinguishable (see
Fig. 9b).

The fact that the Kormendy relation of radio galaxies is not
significantly different from that of normal ellipticals indicates that
the formation processes and the structure of galaxies hosting radio
sources are similar to those of radio-quiet ellipticals.  To go deeper
in to this question, we  obtained spectroscopic observations of a
number of galaxies belonging to our sample, aimed at measuring their
velocity dispersions and at investigating the whole fundamental plane
of radio galaxies.  This analysis is in progress and the results will be
reported  in a forthcoming paper.

\subsection{Colors of the galaxies}

For 29 sources in our sample we obtained both B and R images.
Photometric and structural profiles for these sources have been
reported in Paper I. From these data we derived
the integrated colors $B-R$ and the color profiles as a function of the
semimajor axis. 
Colors ($B-R$) in Table 6 (column 2) have been corrected for
galactic extinction, k-correction and nuclear component. 
The distribution of the $B-R$ color 
is compared in Fig. 10a with that derived from a sample
of 39 radio quiet elliptical galaxies
(Peletier et al. 1990). 
It turns out that, the integrated color distribution of
radio galaxies is much broader and has a blue tail than elliptical galaxies.
In our sample the average integrated color is
$\langle B-R \rangle =1.39\pm 0.18$ compared with
$\langle B-R \rangle =1.57\pm 0.09$ of Peletier et al. (1990).

\begin{figure}
\resizebox{\hsize}{!}{\includegraphics {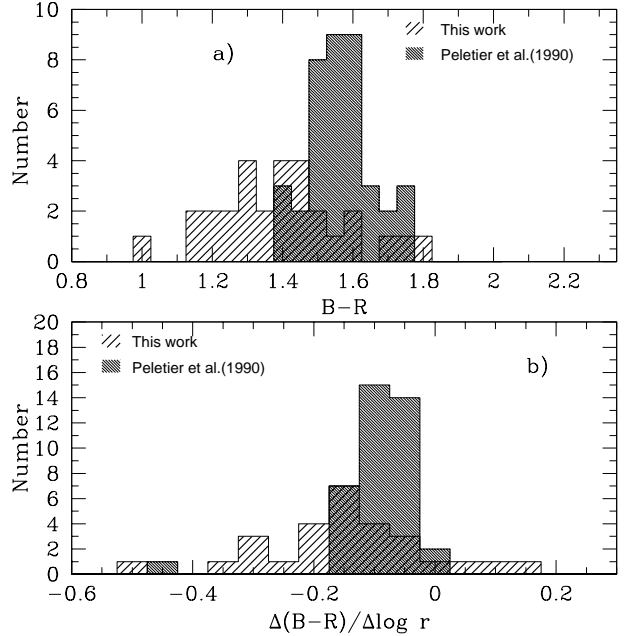}}
\hfill
\caption{a) Distribution of the B-R color of radio galaxies
compared with normal
ellipticals (Peletier et al 1990). b) Distribution of color gradient
of radio galaxies (thinly-hatched)
compared with normal ellipticals (thickly hatched).}
\label{PScolore}
\end{figure}

For each object we have also derived the color profile as a function
of semi-major axis, combining luminosity profiles in the B and R
bands.  The color profiles against the logarithmic radius are well
represented by a linear fit, but in most objects it appears rather
smooth with a systematic tendency of the galaxies to become bluer in
the outer regions.

To quantify this behavior we computed the color gradient
$\Delta(B-R)$/$\Delta log~r$ for each object (see Table 6 column 3) by
a linear regression of the color profile.

The mean color gradient is 
$\langle \Delta(B-R)$/$\Delta log~r\rangle= -0.16\pm 0.17$, compared with 
$\langle \Delta(B-R)$/$\Delta log~r \rangle= -0.09\pm 0.07$ 
for elliptical galaxies of Peletier et
al. (1990).  The average color gradient in radio galaxies appears
therefore slightly steeper than in normal ellipticals while the
distribution of color gradient is much broader (see Fig. 10b).

These results on the integrated colors and color gradients are in
agreement with the results of Zirbel (1996) who found that on average
radio galaxies are bluer by 0.18 mag and have a color dispersion
larger than bright cluster member.  
However, contrary to the finding by Zirbel (1996) we don't find significant
difference in the integrated color distribution between FRI and FRII.
Also Smith \& Heckman (1989) in a
study of multicolor surface photometry of powerful radio galaxies,
found that in general radio galaxies with strong optical emission line
spectra have unusually blue average colors relative to giant
elliptical galaxies while radio galaxies with weak or no emission
lines have colors and color gradients indistinguishable from those of
normal giant ellipticals.
 
The color variation inside each galaxy and among different galaxies is
usually taken to indicate changes in the mean metallicity and/or in
the age of the stellar population (Franx \& Illingworth 1990).
However, the presence of dust could also play a role in the observed
color gradient of galaxies (see e.g. Wise \& Silva 1996).

The color differences observed could be either intrinsic or due to
different dust-reddening for radio galaxies (or a combination of the
two effects).  The fact that the color profiles are rather smooth and
that there is a tendency for the bluer galaxies to have steeper color
gradient (see Fig. 11), support the idea that color differences are mostly
intrinsic.

As a further comparison concerning the colors of radio galaxies 
we show in Fig. 12 the color-magnitude relation of our sample
compared with standard color-magnitudes relations for elliptical galaxies 
(Peletier et al. 1990; Lopez-Cruz 1997). It turns out clearly that 
radio galaxies systematically deviate from the standard
color-magnitude relationship even if some of the brightest 
non-radio ellipticals cover the same region of the radio galaxies.
A possible suggestion is that the color-magnitude relationship
undergoes a breakdown at the highest luminosities.

\begin{figure}
\resizebox{\hsize}{!}{\includegraphics {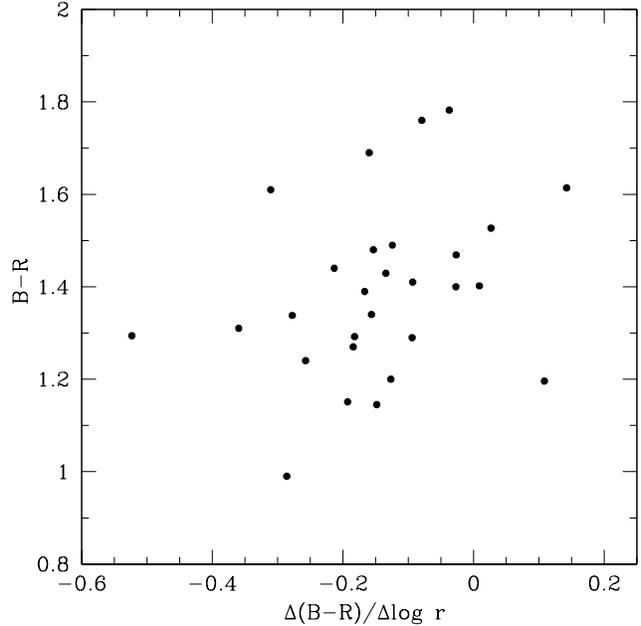}}
\hfill
\caption{Relation between B-R color and color gradient of our radio galaxies}
\label{PScolore}
\end{figure}
\begin{figure}
\resizebox{\hsize}{!}{\includegraphics {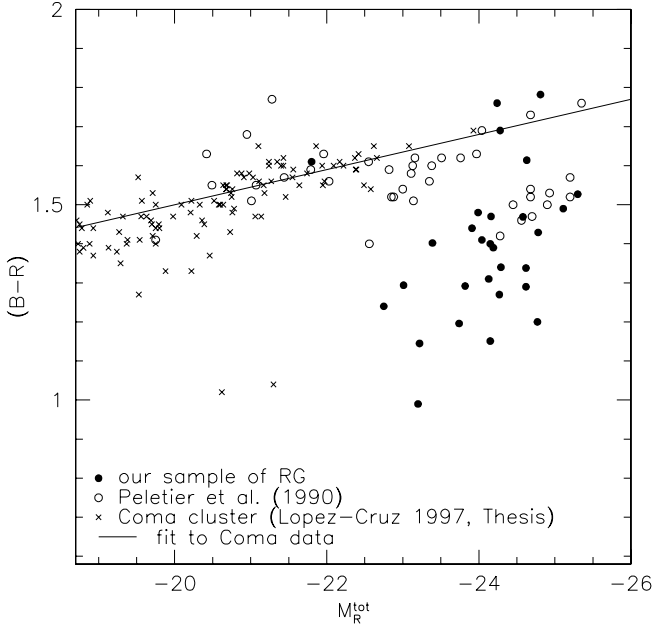}}
\hfill
\caption{Color-magnitude relation of our sample compared with standard
color-magnitude relations for elliptical galaxies (Peletier et al. 1990;
Lopez-Cruz 1997)}
\label{PScolore}
\end{figure}

\section{Structural and morphological properties}

\subsection{Ellipticity}
\begin{figure}
\resizebox{\hsize}{!}{\includegraphics {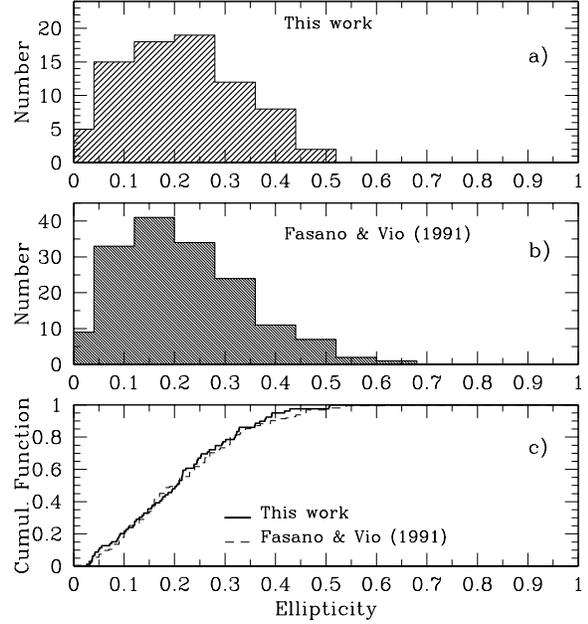}}
\hfill
\caption{a) Distribution of ellipticity calculated at the effective radius. b) Distribution of ellipticity in a sample of non radio ellipticals 
(Fasano \& Vio 1991). Panel c) shows the comparison
of the two cumulative distributions.}
\label{PSell}
\end{figure}
The distribution of the observed ellipticity $\epsilon$ in a sample of
elliptical galaxies gives interesting information about the intrinsic
shapes of the galaxy population. Introducing some "a priori"
hypothesis on the geometrical properties of the galaxy body (oblate,
prolate, triaxial), one can deproject the observed ellipticity
distribution to infer the distribution of the intrinsic
ellipticities. In practice, even if deprojecting ellipticity
distributions is an ill-posed problem (generally it has no unique
solution), a significant difference between the observed distributions
of two galaxy populations implies a difference in their intrinsic
shape.

From surface photometry analysis we obtained for each object the
ellipticity profile.  In order to characterize the ellipticity of the
galaxies we took the values of $\epsilon$ at the effective radius (see
column 9 in Table 5).

From our sample we find
 $\langle \epsilon_e \rangle$=0.21$\pm$0.12,
while the mean ellipticity at $\mu_R=24.5$ is
$\langle \epsilon_{24.5} \rangle$=0.25$\pm$0.12.
Considering the two subsamples of FRI and FRII we did not find any
significant difference (P$_{KS}$=0.52) between
the two populations:\\ 
$ \langle \epsilon_e \rangle_{FRI}$=0.21$\pm 0.11$\\
$ \langle \epsilon_e \rangle_{FRII}$=0.21$\pm 0.12$.   

The distribution of $\epsilon _e$ from our sample is shown in
Fig. 13a.  We compared this distribution with the corresponding one
from a large data-set of normal ellipticals by Fasano \& Vio
(1991)(Fig. 13b).  It is found that both distributions are peaked
around 0.2. The KS test (Fig. 13c) shows that the two data sets are
likely to be draw from the same parent population ($P_{KS}$ = 0.97).

This result is at variance with the suggestion by Disney et al. (1984), and
Calvani et al. (1989), that radio galaxies tend to be rounder than
radio-quiet ellipticals.

Our result confirms the previous finding of Smith \& Heckman (1989b)
that powerful radio galaxies have identical ellipticity distribution
of non radio ellipticals.  They argued that the difference noted by
Disney et al. (1984) may be due to different radio power of the
galaxies considered.

Again, no difference in ellipticity between radio galaxies and 
radio-quiet ellipticals was found by Ledlow \& Owen (1995) comparing 
the distribution of $\epsilon_{24.5}$ from a sample of radio galaxies 
(FRI) in clusters with that relative to a control sample of normal 
galaxies in the same environment.

It is noticeable that even the host galaxies of quasars appear to
exhibit the same ellipticity distribution of normal galaxies (McLure
et al. 1999) supporting the idea that this morphological parameter is
very little (if not at all) influenced by nuclear activity.

\subsection{Isophote shape: disky and boxy}

If the residual intensity variations along a best fitting ellipse are
expanded as a Fourier series in the azimuthal angle $\theta$, the
coefficient $c4$, associated with the $\cos(4\theta)$ term is found to
have a relatively large amplitude.  The amplitude of the $c4$
parameter, indicates deviations of the isophote shape from perfect
ellipse in the form of {\it disky} isophotes ($c4$ $>$ 0) or {\it
boxy} isophotes ($c4$ $<$ 0).  We have examined the radial profile of
$c4$ and evaluated the global amplitude of the $c4$ parameter using
two different methods. In the first one we calculated the value of
$c4$ at the effective radius, in the second one we derived for each
galaxy a weighted average value of $c4$ over the profile.  The two
different methods give similar results. Therefore in the following we
will use the percentage value of c4 at the effective radius (Table 5, column 12).

In Fig. 14a we show the distribution of c4 from our sample of radio
galaxies. We find that about 50\% of our objects are {\it boxy} and
50\% are {\it disky}. Moreover the distributions of $c4$ taken for the
sample of FRI and FRII galaxies are essentially the same.  Finally
there is no correlation between radio power and shape of isophotes
(see Fig. 14b).

This is in contrast with the results by Bender et al. (1987) who found
a correlation between isophote shape and radio emission in elliptical
galaxies in the sense that radio-loud objects have generally {\it
boxy} or irregular isophotes.  On the other hand, two more studies
concerning the morphology of radio galaxies support our
conclusion. First, the fraction of {\it boxy} ellipticals found by
Gonzalez--Serrano et al. (1993) in radio galaxies is significantly
lower than that reported by Bender et al. (1987).  Second, the radio
galaxies in Abell cluster studied by Ledlow \& Owen (1995), do not
show an higher percentage of {\it boxy} isophotes with respect to a control
sample of radio-quiet ellipticals.

\begin{figure}
\resizebox{\hsize}{!}{\includegraphics {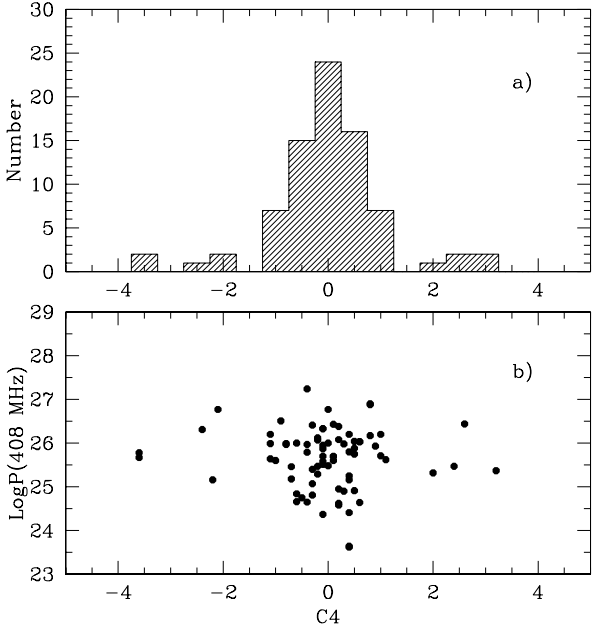}}
\hfill
\caption{a) Histogram of the c4 parameter for our sample of radio galaxies;
b) Radio power at 408 MHz versus the c4 parameter.}
\label{PSc4}
\end{figure}
\section {Signature of interaction}

The isophotal analysis of the host galaxies is important to understand
whether or not there are signatures of gravitational
interaction. These signatures include isophote twisting,
non-concentric isophotes and/or the presence of close companions with
signs of a disturbed morphology.  In this section characteristic
values for these morphological distortions are evaluated and compared
with the corresponding values taken from other samples of both radio
and non radio ellipticals.

\subsection{Non concentric isophotes}

The presence of non concentric isophotes is among the strongest evidences
of interaction between members of galaxy pairs or groups.
We quantify the presence of non concentric isophotes by the parameter
$\delta=\Delta R/R$ where $\Delta R=[(X_c - X_o)^2+(Y_c-Y_o)^2]^{1/2}$.
It represents the percentage of displacement
of the isophotes with respect to the size of the galaxy.
$X_o$ and $Y_o$ represent the position of the center of the inner isophotes,
while $R$ is the radius of the isophote, having $X_c$ and $Y_c$ as center
coordinates.
Since $\delta$ may slightly change with R we took the value at the
effective radius $R_e$.
In Fig. 15 we show the histogram of the $\delta$ parameter.    
The average value of $\delta$ for the whole sample is 0.03 with no
significant difference between galaxies hosting FRI and FRII
radio sources.
\begin{figure}
\resizebox{\hsize}{!}{\includegraphics {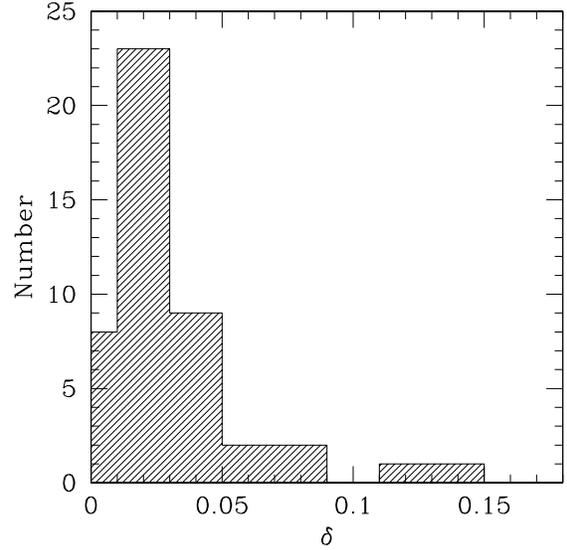}}
\hfill
\caption{Distribution of the $\delta$ parameter}
\label{PSdelisof}
\end{figure}

Apparent isophote displacement could be induced by the photometric
superposition of two galaxies as in the case of Dumbbell systems.
Therefore for objects with large close companions we have adopted a
two recursive fitting analysis (see Paper I).  It turned out that in
most cases no significant displacement of isophotes is present in the
two galaxies.

In a sample of 44 FRI radio galaxies studied by Colina \& De Juan
(1995), the average value of $\delta$ is 0.05. Comparing their result
with those of 40 normal elliptical galaxies studied by Sparks et
al. (1991) ($\langle \delta \rangle=0.017$), they concluded that FRI
host galaxies are more widely distributed over $\delta$ and show much
larger values of $\delta$ than non radio ellipticals.  Our result,
lies between the previous two and suggests again that the role of
interactions for radio galaxies is less relevant than previously
believed.

\begin{table*}
\begin{center}
\begin{tabular}{cccccccccccr} 
\multicolumn{12}{c}{{\bf Table 5.} Photometric and structural parameters of radio galaxies.}\\
\multicolumn{12}{c}{}\\
\hline\\ 
\multicolumn{1}{c}{IAU name} & \multicolumn{1}{c}{$R_e$(kpc)} & \multicolumn{1}{c}{$\mu _e$}& \multicolumn{1}{c}{$m_{24.5}$} & \multicolumn{1}{c}{$m_{tot}$} & \multicolumn{1}{c}{$M_{HOST}(24.5)$} & \multicolumn{1}{c}{$M_{HOST}(tot)$} & \multicolumn{1}{c}{
$M_{PS}$}
  & \multicolumn{1}{c}{$\epsilon _e$} & \multicolumn{1}{c}{$\Delta PA$}
&\multicolumn{1}{c}{$\delta$} & \multicolumn{1}{c}{c4}\\
\medskip (1)     & (2)	& (3) & (4)& (5) & (6)& (7) &(8)&(9) &(10)&(11)&(12) \\   
\hline\\
$0005-199$&   25.2 &      22.51&  15.15 &   14.95&  $-24.38$ &$ -24.58$ &  $ -20.68$   &     0.33  &     3.8    &  0.041    &  $ 0.8$ \\  
$0013-316$&   34.8 &      23.25&  14.98 &   14.62&  $-24.24$ &$ -24.62$ &  $ -20.84$   &     0.21  &    15.7    &  0.029    & $ -0.1$ \\  
$0023-333$&   51.1 &      23.20&  13.16 &   12.88&  $-24.34$ &$ -24.62$ &    ...     &     0.38  &     5.2    &  0.034    &  $-3.6$ \\  
$0034-014$&   10.0 &      21.42&  14.72 &   14.58&  $-23.6$ &$ -23.74$ &  $ -20.29$   &      0.04  &     9.3    & ...      &  $ 0.2$ \\  
$0055-016$&   12.7 &      21.30&  13.22 &   13.11&  $-24.04$ &$ -24.15$ &    ...     &    0.03  &    10.2    &  0.023     &     $  0$ \\  
$0123-016$&   18.9 &      21.68&  11.33 &   11.16&  $-23.87$ &$ -24.04$ &   $<-18.8$ &       0.36  &     5.9    &...      &  $ 3.2$ \\  
$0131-367$&   14.0 &      21.14&  12.29 &   12.18&  $-24.05$ &$ -24.16$ &    ...     &      0.30  &    39.4    &  0.056   &  $ 3.1$ \\  
$0229-208$&   13.7 &      21.59&  14.82 &   14.64&  $-24.01$ &$ -24.19$ &  $<-21.5$  &     0.14  &    ...     &  ...      &  $ 0.3$ \\  
$0247-207$&    32.9&      22.46&  13.63 &   13.43&  $-25.09$ &$  -25.3$ &  $ -21.78$   &     0.12  &    22.5    &  0.039    & $ -0.2$ \\  
$0255+058$&   5.4  &      20.87&  12.92 &   12.75&  $-22.81$ &$ -22.98$ &  $ -18.75$   &    0.03  &    22.8    &  0.015     & $ -0.1$ \\    
$0257-398$&   7.4  &      20.49&  14.48 &   14.38&  $-23.64$ &$ -23.74$ &   $<-21.3$ &     0.21  &    11.9    &  0.012    &   $ 1.0$ \\  
$0307-305$&   9.6  &      21.29&  14.81 &   14.70&  $-23.19$ &$ -23.31$ &  $ -20.87$   &     0.30  &     7.8    &  0.048   &  $ -1.0$ \\  
$0312-343$&    22.5 &      22.57&  14.22 &   14.01& $-23.83$ &$ -24.06$ &  $  -21.3$   &     0.20  &     3.1    &  0.026   & $ -0.2$ \\  
$0325+023$&   20.0 &      22.16&  12.90 &   12.70&  $-23.44$ &$ -23.64$ &   $<-19.1$  &     0.39  &       3    &  0.015   & $ -1.1$ \\  
$0332-391$&   12.1 &      21.49&  14.21 &   14.06&  $-23.72$ &$ -23.88$ & $  -20.93$   &     0.18  &     9.6    &  ...      & $  0.5$ \\  
$0344-345$&   9.0 &      21.78&  14.83 &   14.70&   $-22.7$ &$ -22.84$ &  $ -19.88$   &     0.21  &     5.8    &  0.023   & $ -0.8$ \\  
$0349-278$&   17.8 &      23.03&  15.34 &   15.06&  $-22.72$ &$ -23.01$ & $  -19.77$   &      0.32  &    10.4    & ...   &  $ 2.6$ \\  
$0427-539$&   20.2 &      21.94&  12.82 &   12.69&  $-24.04$ &$ -24.17$ &  $<-20.3$  &    0.33  &     9.8    &   ...   & $ -0.8$ \\  
$0430+052$&   7.4 &      21.25&  13.07 &   13.01&   $-23.06$ &$ -23.15$ &  $ -22.68$   &    0.09  &      13    &  0.038   &  $ 2.4$ \\  
$0434-225$&   27.8 &      22.36&  13.56 &   13.33&  $-24.64$ &$ -24.87$ &  $ -20.38$   &   0.10  &    20.1    &  0.035   &  $ 0.1$ \\  
$0446-206$&   10.1 &      21.79&  14.99 &   14.81&  $-23.36$ &$ -23.54$ &  $<-19.3$  &   0.05  &    16        &0.009   & $ 0.5$ \\  
$0449-175$&   14.5 &      21.88&  12.76 &   12.50&  $-23.65$ &$ -23.91$ &    ...     &   0.17  &     7.8    &    ...   &  $ 0.3$ \\  
$0452-190$&   16.8 &      22.14&  13.26 &   13.00&  $-23.66$ &$ -23.92$ &    ...     &     0.21  &     7.9    &  0.001   & $ -0.4$ \\  
$0453-206$&    19.2 &      22.15&  12.75 &   12.57& $-23.93$ &$ -24.11$ & $<-19.5$  &     0.10  &    17.5    &  0.027   & $ -0.4$ \\  
$0511-305$&   11.4 &      22.30&  14.88 &   14.66&  $-22.94$ &$ -23.16$ &  $<-19.1$  &     0.12  &    21.4    &  ...   &  $ 0.2$ \\  
$0533-377$&    24.1 &      22.83&  14.85 &   14.57& $-24.14$ &$ -24.42$ & $<-20.0$  &    0.12  &    17.7    &  0.019   &  $ 1.1$ \\  
$0546-329$&   18.5 &      21.68&  12.46 &   12.34&  $-24.34$ &$ -24.46$ &    ...     &     0.20  &     3.1    &  0.015   &  $ 0.4$ \\  
$0548-317$&   9.7 &      21.58&  13.57 &   13.42&   $-23$ &$ -23.16$ &   $ -19.6$   &     0.24  &     4.1    &  0.011   &  $-0.3$ \\  
$0620-526$&   28.9 &      21.93&  12.73 &   12.53&  $-24.77$ &$ -24.97$ &  $ -20.81$   &     0.25  &     8.5    &  ...   &  $ 0.5$ \\  
$0625-354$&   19.6 &      21.91&  13.27 &   13.11&  $-24.32$ &$ -24.49$ &  $ -22.15$   &     0.22  &     6.7    &  ...   &  $ 0.6$ \\  
$0625-536$&   37.9 &      22.23&  12.57 &   12.39&  $-25.08$ &$ -25.26$ &  $<-20.3$ &     0.37  &     8.7    &  0.136   &  $ 0.1$ \\  
$0634-205$&   9.8 &      21.05&    14.10 &   13.90& $-23.64$ &$ -23.84$ &  $<-19.4$ &     0.18  &    28.9    &   ...   & $ -0.9$ \\      
$0712-349$&   11.6 &      21.00&    13.15 &   12.98&$-24.04$ &$ -24.21$ & $<-19.5$ &     0.16  &     3.8    &   0.01   & $ -0.3$ \\      
$0718-340$&   11.0 &      21.20&    12.35 &   12.15&$-23.9$ &$  -24.1$ &  $ -18.65$  &     0.12  &    19.8    &  ...   & $ -0.7$ \\      
$0806-103$&   10.0 &      21.05&    15.27 &   15.20&$-24 $&$ -24.07$ &  $ -20.81$  &  0.09 &  32.9    &     ...     &  $ 0.8$ \\      
$0915-118$&   22.0 &      21.91&    13.35 &   13.23&$-24.3$ &$ -24.42$ &   ...     &     0.19  &   8.4    &    ...     & $ -0.4$ \\      
$0940-304$&   9.4 &      20.66&    13.25 &   13.12& $-23.61$ &$ -23.74$ &  $<-20.6$  &     0.42   &   7.2    &  0.023     &  $ 0.6$ \\      
$0945+076$&   6.8 &      21.42&    15.29 &   15.19& $-23.06$ &$  -23.2$ &  $ -22.19$  &     0.04 &   2.8    &  0.015   &     $0.8$ \\      
$1002-320$&   21.1 &      22.38&    14.75 &   14.52&$-24.06$ &$ -24.29$ & $<-19.5$   &     0.20  &  16.2    &  0.024   & $ -0.6$ \\      
$1043-290$&   37.4 &      23.38&    13.68 &   13.22&$-24.22$ &$ -24.68$ & $<-19.5$  &     0.09  &   9.5    &    ...  & $ -2.2$\\       
$1053-282$&   29.2 &      22.40&    13.99 &   13.86&$-23.91$ &$ -24.05$ & $  -20.14$  &     0.51  &   4.4    &    ...  & $ -3.6$\\       
$1056-360$&   9.0 &      21.32&    14.84&    14.64& $-23.41$ &$ -23.61$ &  $<-19.8$ &     0.13  &  52.1    &   0.009  & $ -0.4$\\       
$1107-372$&   12.4 &      20.73&     9.61&     9.53&$-24.3$ &$ -24.38$ &   ...     &     0.26  &   5.7    &   0.006  &   $0.4$\\       
$1123-351$&    24.1 &      22.19&    12.15&    11.92&$-24.33$ &$ -24.56$ & $<-18.3$ &     0.15  &   5.3    &   0.019  &      $ 0$\\       
$1251-122$&   12.91 &      20.85&    10.70&    10.63&$ -24.11$ &$ -24.18$ &   ...     &     0.32  &  6.5    &   0.122  &  $ 0.4$\\       
$1251-289$&   17.9 &      20.77&    12.48&    12.41& $ -25.3$ &$ -25.37$ & $<-19.7$  &     0.05  &  70.3    &   0.086  &  $ 0.4$\\       
$1257-253$&   15.2 &      21.79&    14.17&    13.92& $-23.91$ &$ -24.16$ & $<-19.6$  &     0.25  &   5.1    &   0.004  & $ -0.7$\\       
$1258-321$&   17.3 &      21.61&    10.87&    10.71& $-24.21$ &$ -24.37$ &  $ -18.68$  &     0.22  &   5.4    &   0.016  &  $ 0.2$\\       
$1318-434$&   18.1 &      21.56&    10.09&     9.98& $-24.03$ &$ -24.14$ &   ...     &     0.28  &  13.6    &    ...  & $ -0.3$\\       
$1323-271$&    17.4 &      21.54&    13.22&    13.14&$ -23.97$ &$ -24.05$ &$<-18.2$  &    0.51  &   4.5    &   0.032  & $ -0.1$\\       
$1333-337$&   13.8 &      21.02&    10.05&     9.97& $-24.43$ &$ -24.52$ & $<-17.1$ &     0.09  &  29.2    &   0.004  & $ -0.3$\\       
$1344-241$&   6.2 &      20.34&    12.43&    12.37& $-23.01$ &$ -23.07$  & $<-18.6$  &     0.43   &     2    &   0.018  &  $ 0.2$\\       
$1354-251$&   9.5 &      20.95&    13.44&    13.36& $-23.42$ &$  -23.5$ & $<-18.8$  &     0.39  &   5.8    &   0.011  & $ -0.6$\\
\hline
\end{tabular}
\end{center}
\end{table*}

\begin{table*}
\begin{center}
\begin{tabular}{cccccccccccr} 
\multicolumn{12}{c}{{\bf Table 5 (continued).} Photometric and structural parameters of radio galaxies.}\\
\multicolumn{12}{c}{}\\
\hline\\ 
\multicolumn{1}{c}{IAU name} & \multicolumn{1}{c}{$R_e$(kpc)} & \multicolumn{1}{c}{$\mu _e$}& \multicolumn{1}{c}{$m_{24.5}$} & \multicolumn{1}{c}{$m_{tot}$} & \multicolumn{1}{c}{$M_{HOST}(24.5)$} & \multicolumn{1}{c}{$M_{HOST}(tot)$} & \multicolumn{1}{c}{
$M_{PS}$}
  & \multicolumn{1}{c}{$\epsilon _e$} & \multicolumn{1}{c}{$\Delta PA$}
&\multicolumn{1}{c}{$\delta$} & \multicolumn{1}{c}{c4}\\
\medskip (1)     & (2)	& (3) & (4)& (5) & (6)& (7) &(8)&(9) &(10)&(11)&(12) \\   
\hline\\ 
$1400-337$&    25.6 &      21.86&     9,89&     9.72& $-24.75$ & $-24.92$ &$<-19.0$    &   0.22  &  17.3    &   0.012  &  $-0.5$\\    
$1404-267$&   16.2 &      21.84&    11.84&    11.71&  $-23.8$ & $-23.93$ &  $ -18$     &     0.19  &   6.7    &   0.017  & $ -0.1$\\       
$1514+072$&   41.6 &      22.70&    12.06&    11.81& $-24.56$ &$ -24.81$ &  $ -17.97$  &     0.36  &     3    &   0.038  & $ -0.2$\\       
$1521-300$&   3.6 &      20.83&    13.70&    13.64& $-21.74$ & $ -21.8$ &  $<-19.1$  &     0.24  &  11.6    &   0.061  &  $ 0.4$\\       
$1637-771$&   6.9 &      20.49&    13.64&    13.54& $-23.35$ &$ -23.45$ &  $ -20.15$  &     0.32  &  33.9    &     ...  &  $ 0.9$\\       
$1717-009$&   9.4 &      22.16&    14.07&    13.87& $-22.32$ &$ -22.52$ &  $ -18.39$  &     0.04  &    35    &    0.01  &    $   0$\\       
$1733-565$&   10.8 &      21.47&    15.39&    15.22& $ -23.6$ &$ -23.78$ & $  -20.54$  &    0.21  &    18    &   0.029  &  $-2.1$\\       
$1928-340$&   21.3 &      21.90&    14.44&    14.31& $ -24.6$ & $-24.73$ & $<-21.1$ &     0.14  &     5    &     ...  & $ -1.1$\\       
$1929-397$&   35.7 &      22.57&    13.44&    13.25&$ -24.92$ &$ -25.11$ &   ...     &     0.24  &  23.6    &     ...  &   $ 1.0$\\    
$1949+023$&   9.2 &      20.85&    14.20&    14.11& $-23.66$ &$ -23.75$ & $<-19.2$  &   0.25 & 16.3    &     ...  &  $-0.1$\\  
$1954-552$&   8.0 &      20.94&    14.50&    14.40& $-23.28$ &$ -23.39$ &  $ -19.9$ &   0.24  &  4.8    &    ...  & $ -0.1$\\  
$2013-308$&   15.5 &      21.40&    14.03&    13.95& $-24.69$ &$ -24.78$ &    $ -22$ &   0.17  & 37.6    &    ...  &  $-0.1$\\  
$2031-359$&   23.1 &      22.36&    14.28&    14.14& $-24.48$ &$ -24.63$ & $ -20.75$ &   0.15  &   18    &    ...  &  $ 0.4$\\  
$2040-267$&   9.6 &      20.96&    12.97&    12.91& $-24.0$7 & $-24.13$ & $<-21.1$  &   0.03  & 25.5    &     ...  &  $ 0.1$\\  
$2058-282$&   27.3 &      23.02&    12.94&    12.65& $-23.99$ &$ -24.28$ &   ...   &   0.07  &   11    &    ...  &  $ 0.6$\\  
$2059-311$&   17.8 &      21.58&    12.80&    12.66&$ -24.13$ &$ -24.27$ & $<-19.2$  &   0.38  &  1.5    &    ...  & $ -0.6$\\  
$2104-256$&   23.4 &      22.46&    12.97&    12.63& $ -23.9$ &$ -24.24$ & $<-20.3$  &   0.27  &    8    &    ...  &  $-2.4$\\ 
$2128-388$&   5.4 &      20.26&    12.00&    11.97& $-23.19$ &$ -23.22$ &  $-18.67$ &   0.16  &  4.5    &   0.027  &  $ 0.4$\\  
$2158-380$&    6.4 &      20.58&    13.44&    13.41&$ -23.17$ &$  -23.2$ &  $<-20.2$ &   0.28  & 12.3    &    ...  & $ 2.0$\\  
$2209-255$&   12.8 &      21.16&    13.78&    13.72& $-24.23$ &$ -24.29$ &   $<-21.0$  &   0.32  &  4.6    &    ...  &  $ 0.2$\\  
$2221-023$&   2.4 &      19.28&    14.56&    14.46& $-22.17$ &$ -22.42$ &  $-22.88$ &    0.04  &  7.5    &   0.028  & $ -0.3$\\  
$2225-308$&   3.7 &      20.29&    15.10&    15.00& $-22.65$ &$ -22.75$ &    $<-20.5$  &    0.07  & 12.4    &    ...  & $ -0.2$\\  
$2236-176$&   31.0 &      22.88&    13.76&    13.49& $ -24.5$ &$ -24.77$ &    $<-20.3$  &   0.11  & 34.2    &    ...  & $ -1.1$\\  
$2333-327$&   10.5 &      21.26&    13.79&    13.71& $-23.73$ &$ -23.82$ &   $-20.5$ &   0.15  & 15.5    &   0.037  &  $ 0.5$\\  
$2350-375$&   13.5 &      21.61&    15.38&    15.21& $-23.97$ &$ -24.15$ & $ -21.55$ &   0.29  & 18.5    &    ...  &  $ 0.1$\\  
$2353-184$&   10.5 &      21.31&    14.55&    14.38& $-23.82$ &$ -23.99$ &   ...   &   0.05  & 19.6    &   0.073  & $ -0.1$\\  
\hline
\end{tabular}
\end{center}
\end{table*}

\begin{table}
\begin{center}
\begin{tabular}{ccr} 
\multicolumn{3}{c}{{\bf Table 6.} Color index and color gradients}\\
\multicolumn{3}{c}{}\\
\hline\\ 
\multicolumn{1}{c}{IAU name} & \multicolumn{1}{c}{(B-R)} & \multicolumn{1}{c}{
$\Delta$(B-R)/$\Delta$log~r}\\
\medskip (1)     & (2)	& (3) \\   
\hline\\
$0005-199$&       1.47      &   $ -0.027$\\  
$0013-316$&       1.34      &   $ -0.278$\\  
$0023-333$&       1.29      &   $ -0.094$\\  
$0034-014$&       1.20      &   $  0.108$\\  
$0055-016$&       1.4       &   $ -0.027$\\  
$0123-016$&       1.41      &   $ -0.093$\\  
$0131-367$&       1.47      &   $ -0.677$\\  
$0229-208$&       1.39      &   $ -0.167$\\  
$0247-207$&       1.53      &   $  0.027$\\  
$0349-278$&       1.29      &   $ -0.523$\\  
$0449-175$&       1.44      &   $ -0.213$\\  
$1514+072$&       1.78      &   $ -0.037$\\  
$1521-300$&       1.61      &   $ -0.311$\\  
$1929-397$&       1.49      &   $ -0.124$\\  
$1954-552$&       1.40      &   $  0.009$\\  
$2013-308$&       1.43      &   $ -0.134$\\  
$2031-359$&       1.61      &   $  0.142$\\  
$2040-267$&       1.31      &   $ -0.360$\\  
$2058-282$&       1.69      &   $ -0.160$\\  
$2059-311$&       1.27      &   $ -0.184$\\  
$2104-256$&       1.76      &   $ -0.079$\\  
$2128-388$&       1.14      &   $ -0.148$\\  
$2158-380$&       0.99      &   $ -0.286$\\  
$2209-255$&       1.34      &   $ -0.156$\\  
$2225-308$&       1.24      &   $ -0.257$\\  
$2236-176$&       1.2       &   $ -0.127$\\  
$2333-327$&       1.29      &   $ -0.182$\\  
$2350-375$&       1.15      &   $ -0.193$\\  
$2353-184$&       1.48      &   $ -0.153$\\  
\hline
\end{tabular}
\end{center}
\end{table}

\subsection {Isophotal Twisting}
\begin{figure}
\resizebox{\hsize}{!}{\includegraphics {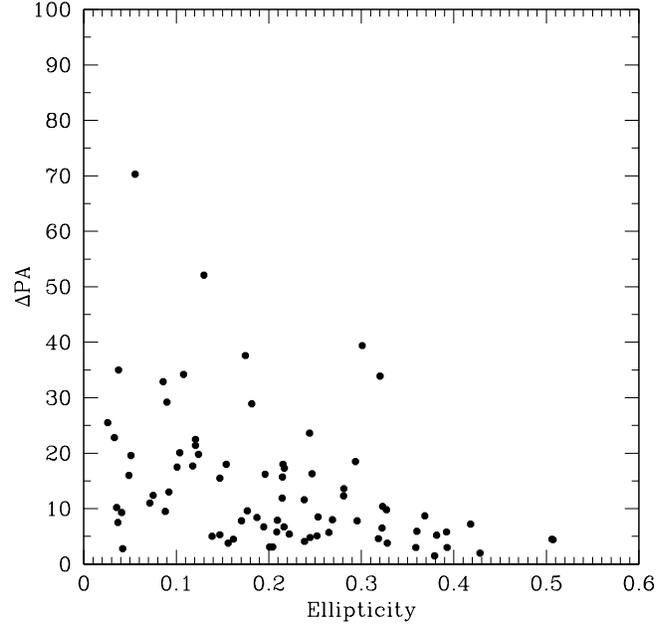}}
\hfill
\caption{Plot of the twisting ($\Delta PA$) versus ellipticity 
at the effective radius for our 
sample of radio galaxies.}
\label{PSellpa}
\end{figure}
\begin{figure}
\resizebox{\hsize}{!}{\includegraphics {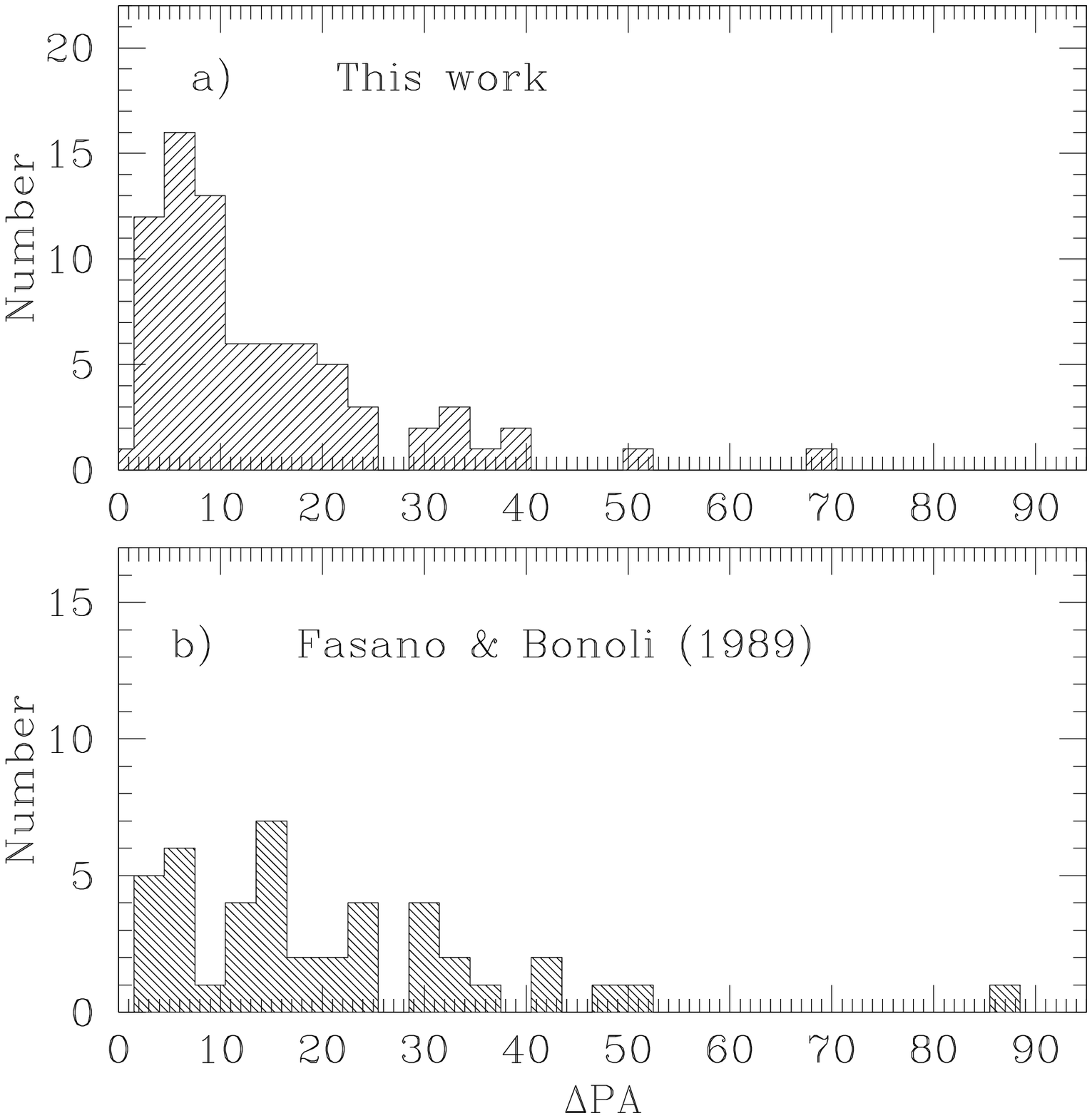}}
\hfill
\caption{a) Distribution of $\Delta PA$ for our sample. b) Distribution of $\Delta PA$ studied by Fasano \& Bonoli (1989) for a sample of 43 isolated elliptical galaxies.}
\label{PStwist}
\end{figure}

Isophotal twisting is indicated by the position angle variation of
the ellipse's major axis as a function of radius.
Kormendy (1982) showed that significant twists are found preferentially
in galaxies with prominent companions rather than in nearly isolated
galaxies, suggesting that tidal effects may account for most large twists
detected to date.
On the other hand, Fasano \& Bonoli (1989) found the amount of isophotal
twisting in their sample of 43 isolated ellipticals to be similar to that
detected in randomly selected samples. They concluded that most
of the twisting observed in elliptical galaxies is intrinsic (triaxiality).

However, it is worth noting that two other factors
may produce apparent isophotal
twisting: the overlapping with isophotes of nearby projected
companions, a residual gradient of the background.
These factors could lead to different results in the position angle
estimates, depending on the different image quality, on the
image processing used and on the criteria used to define a 
twist as significant.
A characteristic of the isophotal twist is that it correlates
with the ellipticity,
in the sense that the roundest galaxies show the largest twists
(Galletta 1980), while very elongated systems show small twists.
 
We give a measure of the isophotal twist in Table 5 (column 10).
The measurement represents the maximum absolute twist over
the range of the observed isophotes.
When computing the maximum absolute twist, because of the seeing
effects, 
we excluded the isophotes having radii less than 5$\arcsec$
and the values of $PA$ having errors larger than $8^{\circ}$.

In Fig. 16 we plot the twisting ($\Delta PA$) versus
the ellipticity at
the effective radius for our sample of radio galaxies.
The typical 
tendency for rounder radio galaxies to show larger twist is confirmed.

In Fig. 17a we plot the distribution of $\Delta PA$ from our sample.
The average value of $\Delta PA$ is $\simeq 14^{\circ}\pm12$ (median $\simeq 10^{\circ}$).
We find no difference in twisting between galaxies hosting different
kinds of radio sources: $\langle \Delta PA(FRI) \rangle=13^{\circ}\pm12$,
$\langle \Delta PA(FRII) \rangle=15^{\circ}\pm12$.

Compared with the finding of Fasano \& Bonoli (1989):
$\Delta PA \simeq 21^{\circ}\pm16$
 (median $\simeq 16^{\circ}$)  for a sample of 43 isolated
elliptical galaxies (see Fig. 17b) we don't find
any indication that radio galaxies exhibit larger isophote twisting
than normal  ellipticals. 
Since we used the same method to derive $\Delta PA$ 
no systematic effects due to different analysis are present.

On the other hand this result contrasts with  
 Colina \& de Juan (1995) that, for a
sample of 44 FRI radio galaxies, found a larger twisting
($\Delta PA \simeq 30^{\circ}$; median $\sim 14^{\circ}$) 
with respect to the Sparks et al. (1991) sample of normal ellipticals
($\Delta PA \simeq 10^{\circ}$; median $\sim 4^{\circ}$).

\section{Summary and conclusions}

We have presented the results of the analysis of R band imaging for 79
low redshift radio galaxies. These were extracted from two flux-limited
samples of radio sources and include objects of both FRI and
FRII morphology in the radio power range from $1.8\times10^{23}$ to
$3.1\times10^{26}$ Watt/Hz.  For
these objects we are able to investigate the optical properties of
the host galaxies and study their relationship with the radio
properties. The main results of this study can be summarized as follows:

1) Galaxies are bulge dominated systems with average absolute magnitude 
$\langle M_{HOST}(tot)\rangle=-24.0$; FRI sources are
hosted in galaxies $\sim$ 0.5 mag more luminous than 
FRII galaxies.

2) Apart from the different average luminosity and size there are
 not other significant structural differences between galaxies
hosting FRI and those hosting FRII radio sources.

3) A substantial fraction ($\sim$ 40\%) of the objects observed show
the presence of nuclear point sources whose luminosity is about few
percent that of the whole galaxy; the luminosity of this component
appears correlated with the core radio power but independent of the
luminosity of the host galaxy.

4) Several objects ($\sim$ 40\%) exhibit deviations of the luminosity
profiles with respect to the r$^{1/4}$ law; these
deviations appear either as lack of light in the inner (r $<$ 2 kpc)
region or as light excess in the outer region (r $>$ 10 kpc)

5) Ellipticity distribution of radio galaxies is indistinguishable from
that of normal ellipticals while isophote twisting and de-centering
of isophotes are comparable with that found in normal
ellipticals.

6) Deviations of isophote from ellipses (disky vs boxy) are in general
rather small and show a homogeneous distribution with no
preference for disky or boxy isophote shape; no correlation is
found between radio power and boxiness of isophotes as suggested by
previous studies.

7) Radio galaxies have bluer integrated colors ($B-R$)  than normal ellipticals
and the color gradients are systematically steeper indicating 
 a slightly bluer population in the outer
region with respect to (non-radio) elliptical. Radio galaxies appear to
deviate significantly from color-magnitude relation found for normal early
type galaxies.

Apart from the  difference in the colors these results indicate that
in general galaxies hosting radio sources have
the same structure and properties as radio-quiet ellipticals.
The main distinguishing feature is the presence of nuclear
 point sources as found in several objects. 
Our results support the idea that all
massive ellipticals may become radio loud at some time and that the
radio activity phenomenon does not change significantly the structural
and photometric properties of the host galaxies. The color differences,
if due to intrinsic differences of stellar population, suggest that 
the star formation rate may have been increased by the activity
phenomenon.

\section*{Acknowledgments} 

This work was partly supported by the Italian Ministry for University and
Research (MURST) under grant Cofin98-02-32.
This research has made use of the NASA/IPAC Extragalactic Database (NED) which is
      operated by the Jet Propulsion Laboratory, California Institute of Technology, under
      contract with the National Aeronautics and Space
Administration.

\end{document}